\documentclass[%
superscriptaddress,
nofootinbib,
 amsmath,amssymb,
aps,
prr,
twocolumn
]{revtex4-1}

\usepackage{graphicx}
\usepackage{dcolumn}
\usepackage{bm}

\usepackage{hyperref} 
\usepackage{xcolor}
\usepackage{float}
\definecolor{cite}{rgb}{0.,0.,0.85}   
\hypersetup{colorlinks,linkcolor={cite},citecolor={cite},urlcolor={cite}}
\usepackage{soul}

\usepackage{lipsum}

\begin{document}


\title{Investigating resonant low-energy electron attachment to formamide: dynamics of model peptide bond dissociation and other fragmentation channels}

\author{Guglielmo Panelli}
\affiliation{Department of Physics, University of Nevada, Reno, Nevada 89557, USA
}%

\author{Ali Moradmand}
\affiliation{%
Department of Sciences and Mathematics, California Maritime Academy, Vallejo, California 94590, USA
}%

\author{Brandon Griffin}
\affiliation{%
 Department of Physics, University of Nevada, Reno, Nevada 89557, USA
}%

\author{Kyle Swanson}
\affiliation{%
 Department of Physics, University of Nevada, Reno, Nevada 89557, USA
}%

\author{Thorsten Weber}
\affiliation{%
 Chemical Sciences Division, Lawrence Berkeley National Laboratory, Berkeley, CA 94720, USA}%

\author{Thomas N. Rescigno}

\affiliation{%
 Chemical Sciences Division, Lawrence Berkeley National Laboratory, Berkeley, CA 94720, USA}%
 
\author{C. William McCurdy}
\affiliation{%
Department of  Chemistry, University of California, Davis,  California 95616, USA}%

\affiliation{%
 Chemical Sciences Division, Lawrence Berkeley National Laboratory, Berkeley, CA 94720, USA}%
 
\author{Daniel S. Slaughter}
\email[Corresponding author: ]{dsslaughter@lbl.gov}
\affiliation{%
 Chemical Sciences Division, Lawrence Berkeley National Laboratory, Berkeley, CA 94720, USA}%

\author{Joshua B. Williams}
\email[Corresponding author: ]{jbwilliams@unr.edu}
\affiliation{%
 Department of Physics, University of Nevada, Reno, Nevada 89557, USA
}%
\date{\today}

\begin{abstract}

We report experimental results on three-dimensional momentum imaging measurements of anions generated via dissociative electron attachment to gaseous formamide. 
From the momentum images, we analyze the angular and kinetic energy distributions for NH$_2^{-}$, O$^{-}$, and H$^{-}$ fragments and discuss the possible electron attachment and dissociation mechanisms for multiple resonances for two ranges of incident electron energies, from 5.3~eV to 6.8~eV, and from 10.0~eV to 11.5~eV. {\it Ab initio} theoretical results for the angular distributions of the NH$_2^{-}$ anion for $\sim$6~eV incident electrons, when compared with the experimental results, strongly suggest that one of the two resonances producing this fragment is a $^2$A$''$ Feshbach resonance.
\end{abstract}

\maketitle

\section{\label{sec:intro}Introduction}
The electron-molecule collision process in which a molecule captures a low-energy electron (i.e., with energy up to $\sim$20~eV), forms a short-lived, unstable molecular anion, and thereafter dissociates into several fragments (one negative ion and all other neutral) is a long-studied process known as dissociative electron attachment (DEA). DEA is among the fundamental electron-molecule collision-based interactions~\cite{Schultz1960,Dorman1966,Comton1967,Melton1972,Trahmar1974,Klots1978,Jungen1979,Curtis1992} and has been found to play an important role in a variety of fields from condensed matter~\cite{Eck2000,Mendes2004,Dulub2007,Walz2010} and gaseous electronics~\cite{Haxton2011} to low-energy plasmas~\cite{MORGAN2000}. 
The low-energy electrons involved in DEA to molecules in natural settings are typically produced as by-products of primary interactions between matter and high-energy photons or particles. 
It has been shown that these electrons play a pivotal role in biological processes such as the triggering of DNA strand breaking and other DNA dissociation processes~\cite{Bouda2000,Pan2003,Abdoul2004,Ptasi2007} and radiation damage of proteins~\cite{Sanche2002}. 

Formamide (HCONH$_2$) is widely considered an archetypal model molecule for the investigation of protein and peptide chemistry due to its simple yet rich structure which includes an amide bond.
The decomposition of formamide into other notable simple organic molecules (e.g., CH, HCN, HCNO, etc.) has been widely studied in experimental and theoretical settings.
Formamide is comprised of many of the progenitors of complex biological molecules such as proteins and nucleic acids and is considered an important link in the evolution of simple biomolecules into complex structures. 
Moreover, formamide has received ample interest due to its N-C amide bond. 
This feature makes formamide a prototypical molecule for the study of electron-capture-induced peptide bond breaking. 
Investigation of low-energy electron disruption of peptide bonds - the links between amino acids in proteins - is necessary for a more complete understanding of protein stability. 

Electron scattering from formamide has been extensively studied previously over a broad range of energies~\cite{Gingell1997,bettegaCollisionsLowenergyElectrons2010,vinodkumarElectronImpactRotationally2014,homemElectronCollisionsAmmonia2014a}. Studies of the dissociation pathways of formamide irradiated by (vacuum) ultraviolet light~\cite{dawleyRadiationProcessingFormamide2014a,barksGuanineAdenineHypoxanthine2010b} or higher energy radiation~\cite{saladinoMeteoritecatalyzedSynthesesNucleosides2015,sivaramanElectronImpactDissociation2014,ferusLaserSparkFormamide2011} have revealed chemical products of biological and technological relevance. Several of these reactions are expected to involve dissociative attachment by low energy secondary electrons. DEA to gaseous formamide has been studied previously, both theoretically~\cite{goumansDissociativeElectronAttachment2009,seydouElectronAttachmentStrongly2005,Li2019} and experimentally~\cite{Hamann2011,szymanskaDissociativeElectronAttachment2012,Li2019}.
Thus far, experimental probes of DEA to formamide have focused on fragment yields~\cite{Hamann2011}. Anion fragment momentum imaging provides further details on the dissociation dynamics for fragments resulting from DEA to formamide.
The work of Hamann et al.~\cite{Hamann2011} investigated DEA to formamide in the energy range of 0--18~eV, and the authors have identified several predominant resonances, of which we probe the major resonances between 5.3--6.8~eV and 10.0--11.5~eV.

Here we utilize three-dimensional (3D) ion momentum imaging with an effusive gas target to examine DEA to formamide. Our focus is on the production of NH$_2^{-}$, O$^{-}$, and H$^{-}$. 
All previous work indicates that the NH$_2^{-}$ yield peak between 5--7~eV consists of two superimposed resonance bands centered at 5.9~eV and 6.8~eV~\cite{Hamann2011,Li2019}. 
Further, the work of Li et al.~\cite{Li2019} suggests that one of these bands is due to a core-excited dipole-supported resonance formed by low-energy electrons with incident energy within 5--7\,eV. The recent Comment of Fedor~\cite{fedorCommentDipoleSupportedElectronic2020} on that work argues that the more common mechanism of involving doubly-excited Feshbach resonances should not be ruled out.  
In the present work, we investigate whether the angular dissociation distributions and fragment kinetic energy distributions of NH$_2^{-}$ exhibit disparities in the 5--7~eV incident electron energy range that may allow a more direct identification of the resonances responsible for DEA. To aid in the analysis, we employ {\em ab initio} theory to identify doubly-excited states in the 5.3--6.8~eV energy range that might lead to NH$ _2^{-}$ production and carry out scattering calculations to determine the expected ion angular distributions, which are compared with measured distributions to confirm the resonant states and product assignments. In addition to an investigation of NH$_2^{-}$ production, this work probes the production of O$^{-}$ and H$^{-}$ via DEA to formamide from incident electrons with energy between 10.0-11.5~eV. We provide feasible production mechanisms along with the results of 3D momentum imaging.

The structure of this paper is as follows. 
We first provide a description of the experimental apparatus in Sec.~\ref{sec:apparatus}, followed by an overview of the offline analysis procedure in Sec.~\ref{sec:analysis}.  In Sec.~\ref{sec:theory}, we give a description of the theoretical methods employed. In Sec.~\ref{sec:results} we propose electron attachment mechanisms and summarize our results for each fragment of interest and provide the momentum images, kinetic energy release estimates, and angular dissociation distributions as well as a brief discussion of the results. 
Lastly, Sec.~\ref{sec:conclusions} contains concluding remarks regarding our results and future directions.

\section{\label{sec:experiment}Experiment}
\subsection{Apparatus}\label{sec:apparatus}
Details on the experimental apparatus can be found in Ref.~\cite{Adaniya2012}. 
Here we provide a brief discussion of the most relevant components and specifications. 
The creation of an effusive formamide molecular target was achieved by a 20~mm-long, 0.3~mm-diameter capillary. 
We limited condensation of the target gas by precisely controlling the temperature of the liquid reservoir (70$^\circ$C), gas tubing (87$^\circ$C), and capillary (111$^\circ$C).
The limitation of residual water in the interaction region was achieved with the use of a liquid-nitrogen cold-trap. The background vacuum, without formamide was below 1$\times$10$^{-8}$~Torr, and during operation we achieved a vacuum level of better than 1$\times$10$^{-6}$~Torr. 

A pulsed electron beam was generated by a commercially-acquired electron gun 
and is directed perpendicular in relation to the capillary, which produces the molecular target.
The device produced 80\,ns pulses (including a $\sim$10~ns rise/fall time) at 50\,kHz repetition rate with a full width at half-maximum (FWHM) of 0.8~eV for the electron energy distribution. 
The electron beam pulse is collimated into a $\sim$1~mm-diameter region in a uniform magnetic field of $\sim$25~G produced by a pair of 0.75~m-diameter Helmholtz coils.

After a delay (90-200~ns, depending on the electron beam energy), the electron pulses were followed by a pulsed electric field in the extraction region of the spectrometer. The spectrometer extracts anions from the electron-molecule interaction region in the direction perpendicular to the electron beam and parallel to the gas jet. 
Anions produced by DEA were accelerated towards a time-sensitive 80~mm multi channel plate (MCP) detector equipped with a  position-sensitive delay-line anode.
The time between the electron pulse and ion fragment contact with the MCP detector was recorded together with the position data in list-mode format, allowing for a thorough offline analysis including cleaning and calibrating. 
The initial momentum of each ion fragment was determined entirely from the detector position and timing data using the equations of motion in the known electrostatic field. 

\subsection{Analysis technique and calibration}\label{sec:analysis}

The anion fragment initial momentum data was generated through offline analysis. The fragment dissociation direction and kinetic energy distributions were determined from the measured time and position coordinates. 
Since all anions of interest in this experiment are of the same charge, anions of different mass are distinguished by their recorded time-of-flight to the detector, which is proportional to the square-root of the singly-charged anion mass.
The time-of-flight also encodes one coordinate of momentum, limiting the mass-resolution of the system. This technique does not allow O$^-$ and NH$_2^-$ to be distinguished by their time-of-flight, therefore we rely on previous work~\cite{Hamann2011} to identify the dominant fragmentation channel at each incident electron energy. Momentum calibration was performed by measurements of H$^-$ and O$^-$ anions from DEA to H$_2$O~\cite{Adaniya2012,adaniyaImagingMolecularDynamics2009,haxtonObservationDynamicsLeading2011}. 

For each anion fragment of interest we determine the ion dissociation distribution in 3D momentum space. 
To display the distribution on two axes with a constant solid angle, we display a 
conical slice of the 3D momentum sphere. 
The kinetic energy and angular distributions are derived from the unsliced momentum. 
Note, that in the analysis we exploited the symmetry about the electron beam direction axis for the electron attachment and subsequent anion dissociation dynamics. Additionally, the error bars shown in this work represent one standard deviation in the Poisson statistical uncertainty.

\section{Theory}\label{sec:theory}
The geometry of ground-state formamide is planar and has C$_s$ symmetry. It is nominally described by the configuration (1-9a$'$)$^2$(1a$''$)$^2$(10a$'$)$^2$(2a$''$)$^2$. Doubly-excited (Feshbach) resonance states can be formed when a colliding electron promotes a target electron in an occupied orbital to an unoccupied orbital, and the colliding electron is captured into the same orbital. The lowest unoccupied molecular orbital (LUMO) of formamide is 3a$''$, which we denote here as $\pi^*$. This is a compact and anti-bonding valence orbital, which is responsible for the shape resonance seen in low-energy elastic scattering near 2.5 eV. By analogy with other systems we have studied, including our recent study of DEA to formic acid~\cite{formic}, we would not expect the low-lying electronic states involving excitation into the $\pi^*$ orbital to serve as parents of a doubly-excited Feshbach resonance, which are more likely to involve double occupancy of a $\sigma^*$ orbital with substantial Rydberg character.

We employ standard electronic structure methods to compute the energies of the relevant neutral and anion states. Some care is needed to obtain a balanced description of a negative ion resonance relative to its parent neutral state which can be sensitive
to the choice of molecular orbitals employed. We have found that state-averaged multi-conguration self-consistent field (MCSCF) orbitals based on the (triplet) excited neutral states, which are parents of the resonance anion states, form a good basis for characterizing the resonances as well as the excited target states. The orbital basis for the target states and scattering calculations were obtained from a state-averaged, complete active space MCSCF calculation on the neutral target, averaging over the four lowest triplet states. We used a triple-zeta basis of Gaussian functions, augmented with an additional s- and p-type function on the oxygen, carbon and nitrogen atoms. Nine orbitals were constrained to double occupancy, and the remaining six electrons were distributed over an active space of three a$'$ and three a$''$ orbitals, which results in 182 configurations for the target states in both A$'$ and A$''$ symmetry. The energies of the five lowest target states are listed in Table~\ref{table:energies}.
\begin{table}
\centering
\begin{tabular}{  c  c  c  c} 
 \hline
 State & Principal Configuration &Energy (eV)&Expt.\\
 \hline
 1$^1$A$'$&(1a$''$)$^2$(10a$'$)$^2$(2a$''$)$^2$& 0.0&\\
 1$^3$A$''$&(1a$''$)$^2$(10a$'$)$^1$(2a$''$)$^2$(3a$''$)$^1$& 4.89&$\sim$6\\
 1$^1$A$''$&(1a$''$)$^2$(10a$'$)$^1$(2a$''$)$^2$(3a$''$)$^1$& 5.18&5.82\\
 1$^3$A$'$&(1a$''$)$^2$(10a$'$)$^2$(2a$''$)$^1$(3a$''$)$^1$&5.46&5.2\\
 2$^3$A$''$&(1a$''$)$^2$(10a$'$)$^2$(2a$''$)$^1$(11a$'$)$^1$&5.74&\\
 \hline
\end{tabular}
\caption{Formamide target states and energies used in scattering calculations; experimental energies from ref.~\cite{Gingell1997}}
\label{table:energies}
\end{table}

The trial wave function for the scattering calculations used here takes the form~\cite{rlm95}
 \begin{eqnarray}
 \Psi^-_{\Gamma_ol_om_o}&=&\sum_{\Gamma}\hat{A}(\chi_{\Gamma}F^-_{\Gamma\Gamma_o})+\sum_id^{\Gamma_o}_i\Theta_i\nonumber \\
 &\equiv &P \Psi + Q \Psi \,.
\label{eq:Psi}
\end{eqnarray}
The first sum contains the direct product of the five {N-electron} neutral target states $\chi_{\Gamma}$ described above and corresponding continuum {orbitals} $F^-_{\Gamma\Gamma_o}${,}  and the second sum runs over {(N+1)}-electron configuration-state functions (CSFs) $\Theta_i$, constructed from bound molecular orbitals. The operator $\hat{A}$ {antisymmetrizes} the {product of} continuum and target wave functions. The functions $\Theta_i$ included in the second sum are of two types. The first type consists of all CSFs that can be constructed consistent with symmetry from the molecular orbitals used to expand the target state functions. This group of CSFs is necessary to relax strong orthogonality constraints between target and continuum functions and to describe short-range correlation effects. The second group of functions $\Theta_i$ includes the complement of $P$-space, i.e. the products of virtual molecular orbitals and the remaining target states (182-5) that are energetically closed. This group of terms is essential in describing target relaxation in the presence of an additional electron. Without such terms the resonance state can appear above rather than below its parent neutral state, and thereby incorrectly appear to be a core-excited shape resonance instead of a narrow Feshbach resonance.

Resonance parameters are obtained from multi-state close-coupling calculations using the well-established complex Kohn method, which has been described previously~\cite{rlm95}. The eigenphase sums are fit to a Breit-Wigner form. We use the computed body-frame S-matrix elements to connect the theoretical results to laboratory-frame angular distributions by computing the so-called entrance amplitude, which is a complex-valued matrix element of the electronic Hamiltonian between the resonance wave function and a background scattering wave function:
\begin {equation}
V(\theta,\phi;\Xi)=<Q\Psi|H_{el}|P\Psi>\, ,
\end{equation}
where $\theta$ and $\phi$ are the polar and azimuthal angles of the electron momentum vector incident on the fixed-in-space molecular target, and 
$\Xi$ labels the internal {nuclear} coordinates of the molecule. When the relative orientation of the fragments is not observed, as is the case here, the angular distribution of the DEA product ions is given by
\begin{equation}
\label{axial-recoil}
\frac{d\sigma_{\bf{DEA}}}{d\theta}\propto \int d\phi |V(\theta,\phi;\Xi)|^2\, .
\end{equation}
The procedures we use to construct the entrance amplitudes from the fitted S-matrix elements have been described in detail elsewhere~\cite{topical,formic} and will not be repeated here.

Our calculations reveal two Feshbach resonances at 5.52 and 5.65~eV of  $^2$A$''$ and $^2$A$'$ symmetry, respectively, whose principal configurations are (...)(2a$''$)$^1$ ($\sigma^*$)$^2$ and (...)(10a$'$)$^1$ ($\sigma^*$)$^2$. The computed entrance amplitudes for these resonances give an expected NH$_2^-$ angular distribution that is compared with the measured distributions below.

\section{Results and Discussion}\label{sec:results}
\subsection{NH$_2^{-}$ resonances at 5.3--6.8~eV}

We rely on the high mass resolution results from Ref.~\cite{Hamann2011}, which is able to distinguish between NH$_2^{-}$ (mass 16.019~u) and O$^{-}$ (mass 15.995~u). This measurement shows that the two fragments occurred at different incident electron energies. The incident electron energies between 5.3--6.8~eV are predominantly NH$_2^-$, while O$^-$ is dominant between 10.0--11.5~eV.

\paragraph{Pathways to formation.}
Here we provide plausible formation mechanisms for NH$_2^-$ and predict the energy threshold for each mechanism (the same is done for H$^-$ and O$^-$ in the following sections). In doing so, we rely on literature values for standard heat of formation ($\Delta H_f^0$), electron affinity (EA), and bond dissociation energy ($D$) for each fragment and bond involved in the processes considered here. These values, along with the corresponding references, are provided in Table~\ref{tab:tab1}.

\begin{table}[t!]
\caption{\label{tab:tab1} Summary of thermodynamic data for formamide fragments and bonds relevant to this work.}
\begin{tabular}{p{1.25in} p{0.75in} p{1.25in}}
        \hline
        \hline
        Compound & $\Delta H_f^0$ (eV) & Electron Affinity (eV)   \\ \hline
        HCONH$_2$ & -1.93~\cite{Bauder1958} & ---\\
        H$_2$CNH & \,\,1.14~\cite{Grela1988}  & \\
        CH$_3$N & \,\,3.27~\cite{anlgov}  & \\
        HCN & \,\,1.4\,\,\,~\cite{Chase1998}  & \\
        CNH & \,\,1.99~\cite{anlgov} & \\
        CO & -1.15~\cite{Chase1998} &  \\
        NH$_2$ & \,\,1.97~\cite{Chase1998} & 0.77~\cite{Wickham1989} \\
        H & \,\,2.26~\cite{Chase1998} & 0.75~\cite{Lykke1991} \\
        O & \,\,2.56~\cite{Cox1989} & 1.46~\cite{Cavanagh2007} \\
        \hline
        \hline
        Bond (compound) & $D$ (eV) & \\ \hline
        C-N (HCONH$_2$) & \,\,4.37~\cite{Lou2007} & \\
        N-H (HCONH$_2$) & \,\,4.71~\cite{Lou2007} & \\
        C-H (H$_3$CCHO) & \,\,3.9\,\,\,~\cite{Lou2007} & \\
        C-O (CH$_2$O) & \,\,7.7\,\,\,~\cite{ruscic2015} & \\
        \hline
        \hline
\end{tabular}
\end{table}

One candidate mechanism for the production of NH$_2^{-}$ from DEA to formamide is through the simple cleaving of the C-N amide bond:
$$
    e^{-} + \textrm{HCONH}_2 \to (\textrm{HCONH}_2)^{\ast -} \to \textrm{HCO} + \textrm{NH}_2^{-} \, .
$$
The thermodynamic values in this process (EA(NH$_2^{-}$) = 0.77~eV and $D$(C-N) = 4.37~eV) indicate a threshold energy of 3.6~eV. This is well below the incident electron energy for the measured resonances. Another possible mechanism for the production of NH$_2^{-}$ anions comes from the fragmentation of HCO via the aforementioned mechanism into neutral H and CO fragments:
$$
    e^{-} + \textrm{HCONH}_2 \to (\textrm{HCONH}_2)^{\ast -} \to \textrm{H} + \textrm{CO} + \textrm{NH}_2^{-} \, .
$$
The thermodynamic threshold for this process is 4.2~eV, given the EA of NH$_2$ along with the standard heat of formation values of formamide, H, CO, and NH$_2$ from Table~\ref{tab:tab1}, which is also well below the incident electron energy for the observed resonance.

\begin{figure}[t!]
\begin{tabular}{c c}
\multicolumn{2}{c}{(a) 5.3\,eV } \\ \hline 
\includegraphics[width=0.48\linewidth]{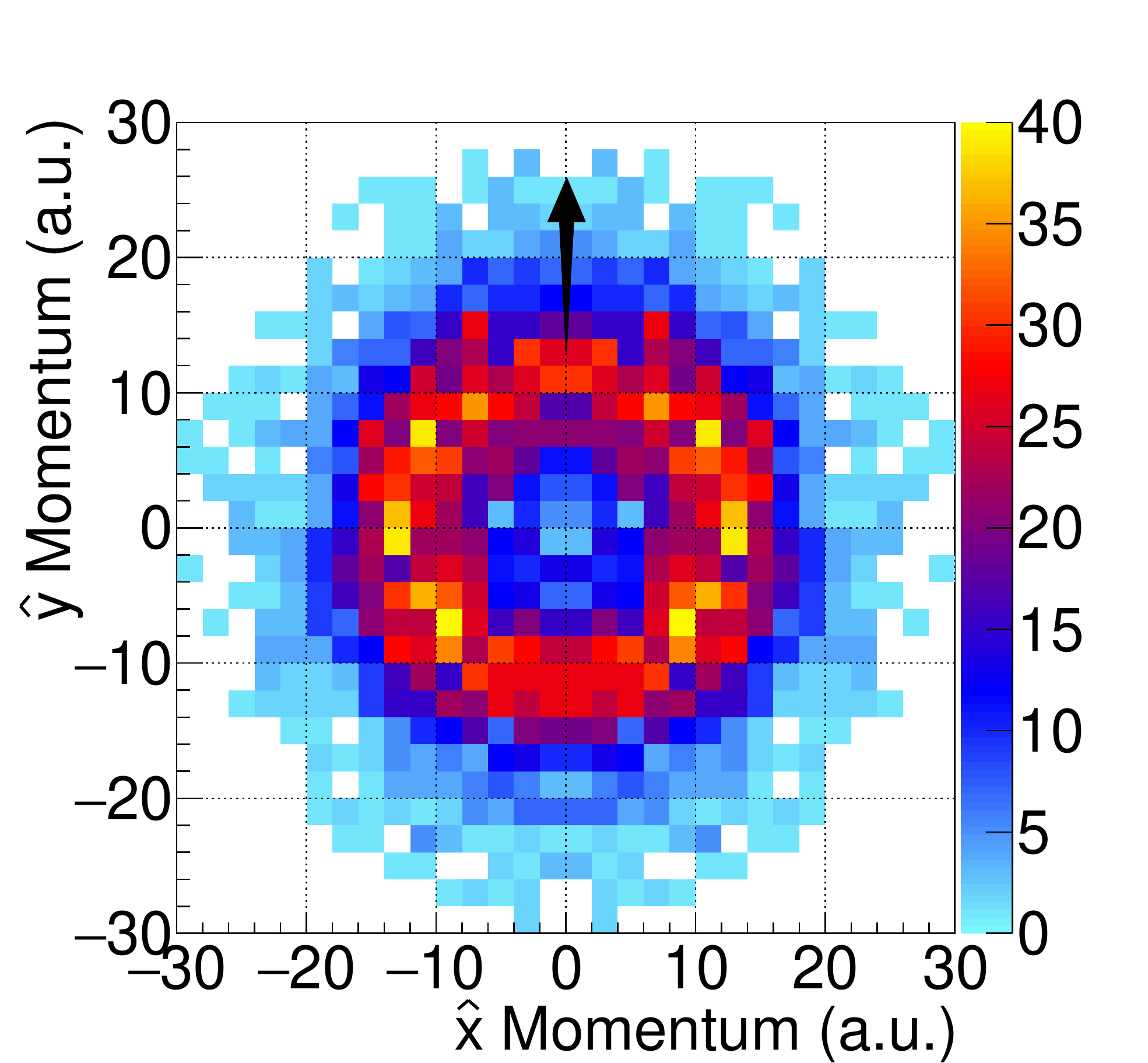} &
\includegraphics[width=0.46\linewidth]{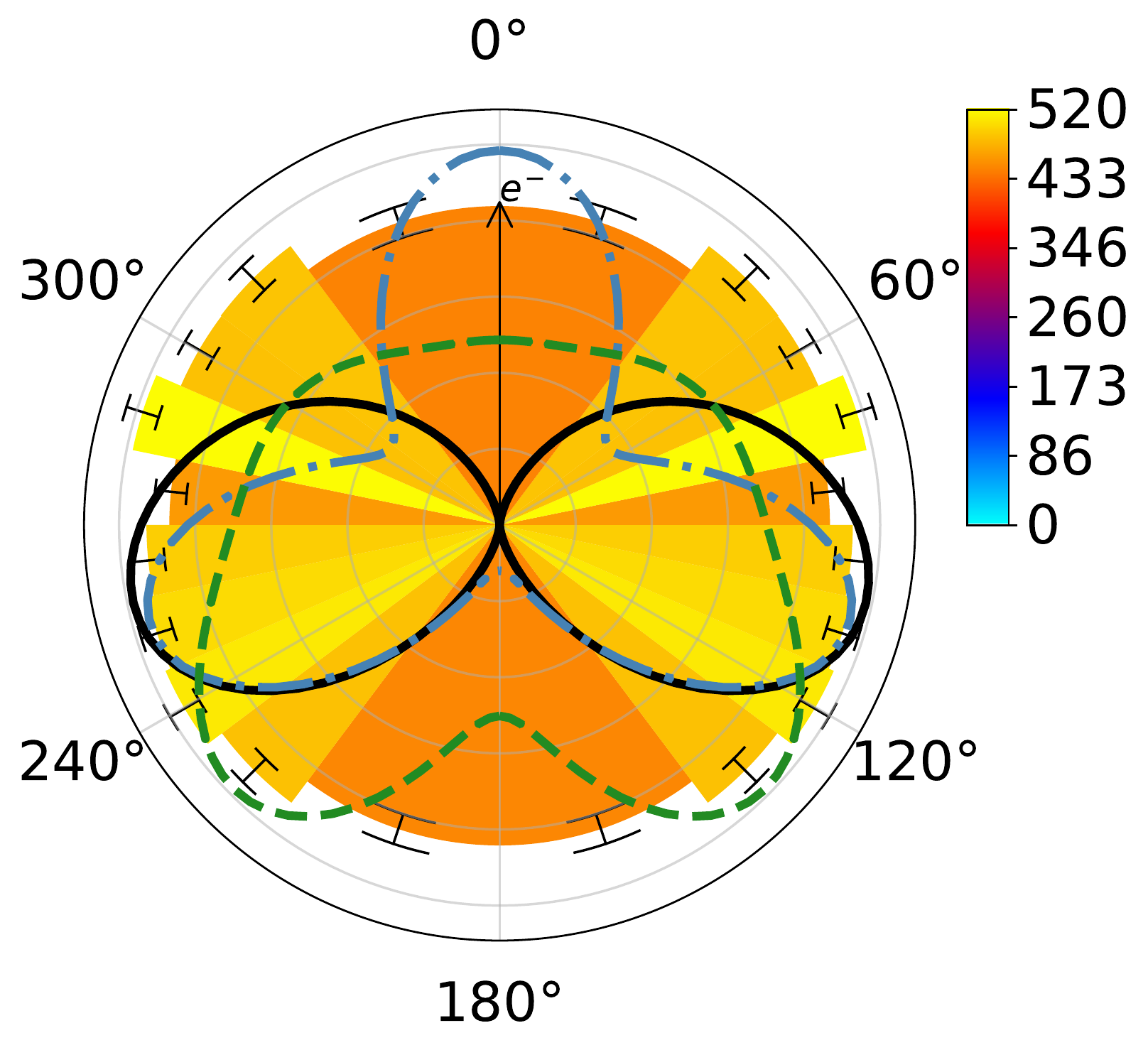} \\  \multicolumn{2}{c}{(b) 5.8\,eV} \\ \hline
\includegraphics[width=0.48\linewidth]{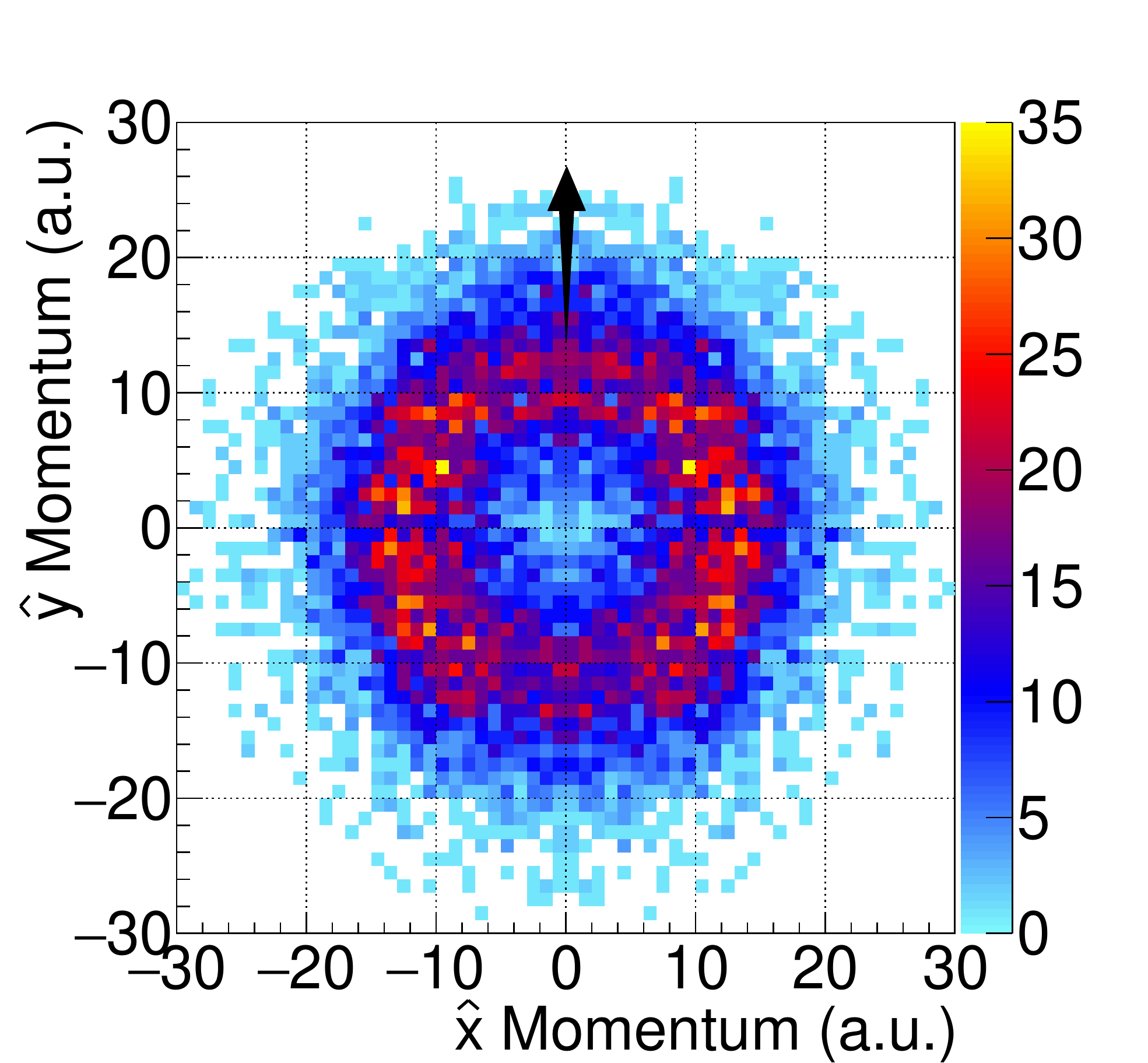} &
\includegraphics[width=0.46\linewidth]{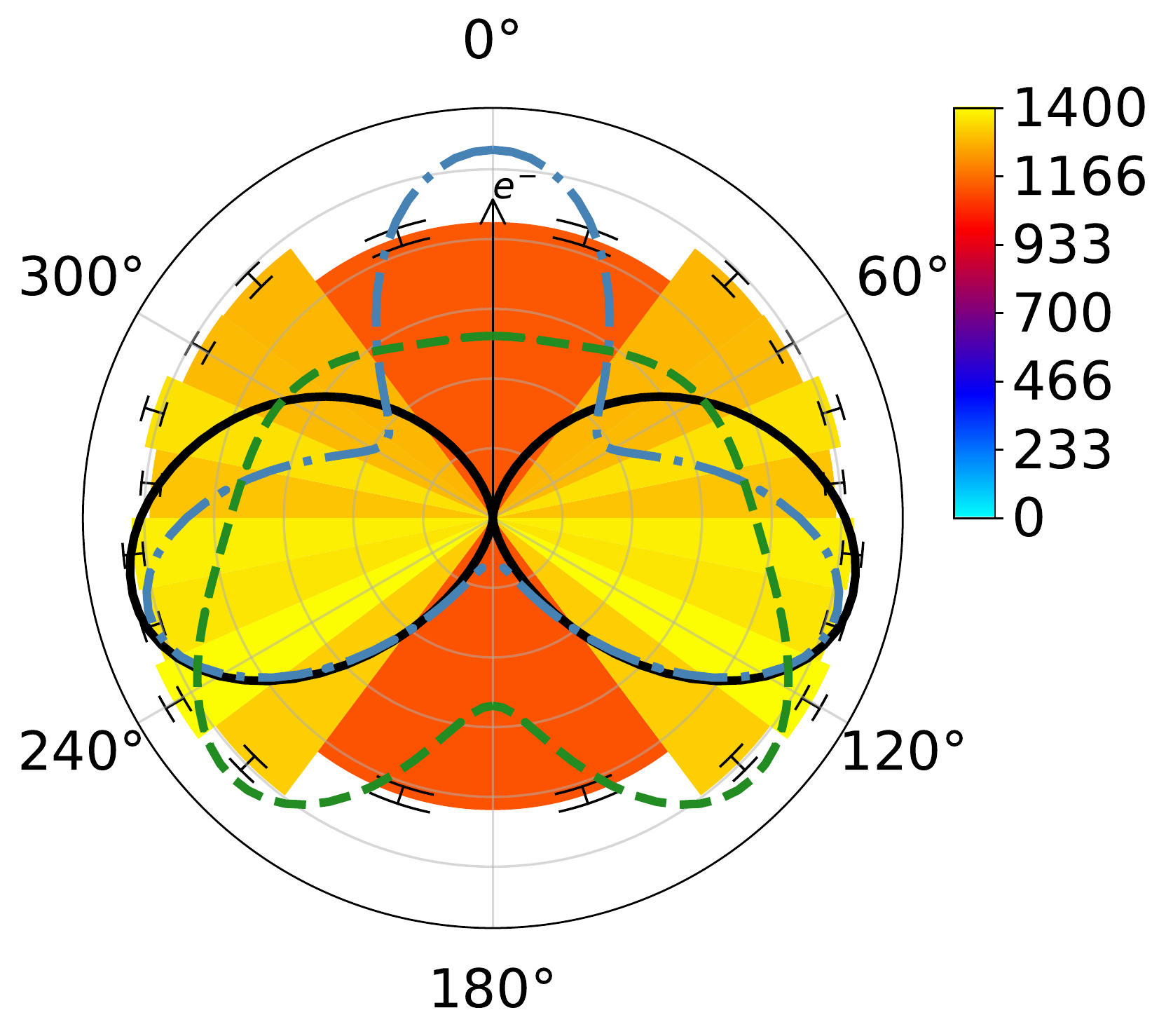}\\   \multicolumn{2}{c}{(c) 6.3\,eV} \\ \hline
\includegraphics[width=0.48\linewidth]{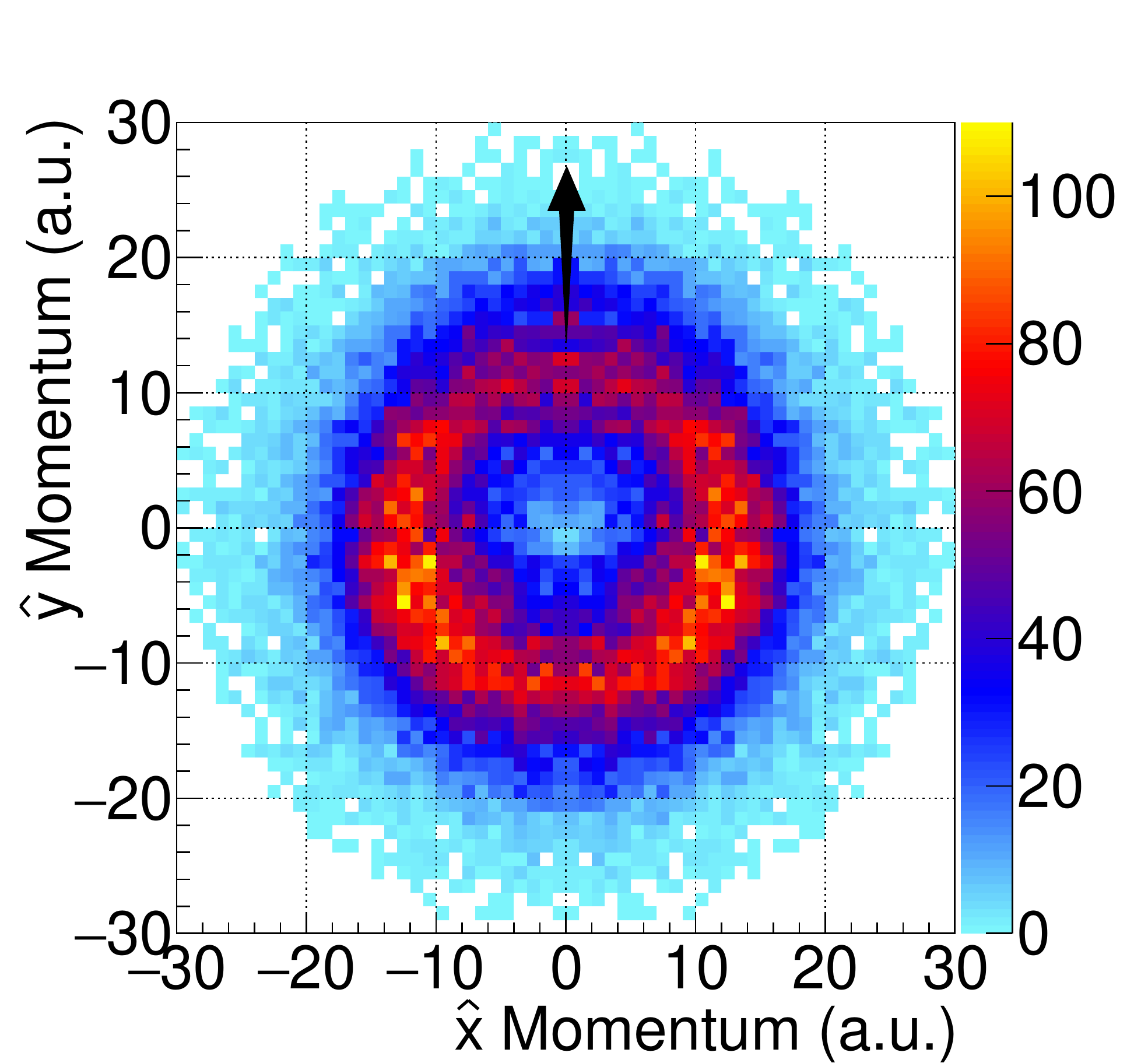} &
\includegraphics[width=0.46\linewidth]{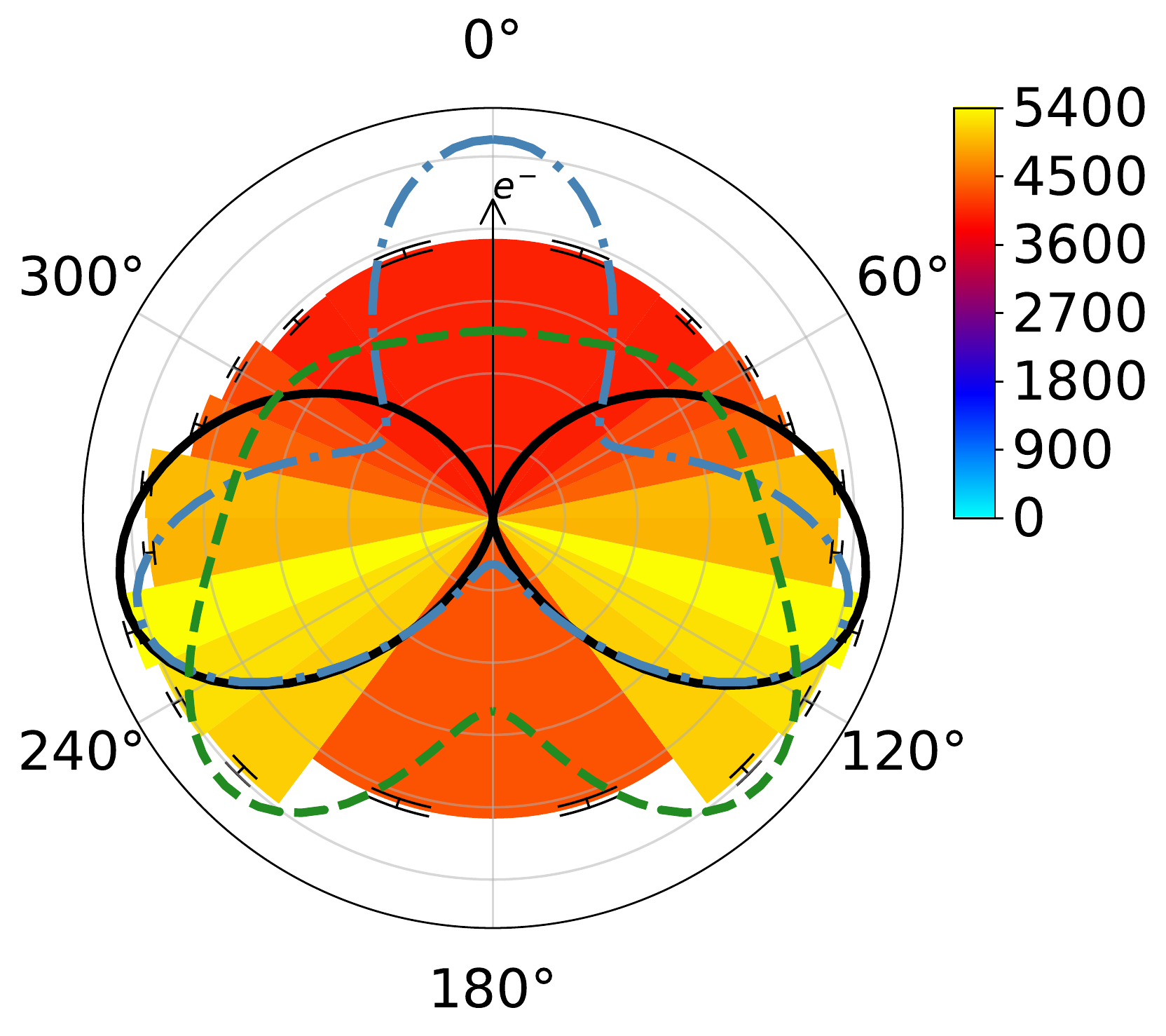} \\  \multicolumn{2}{c}{(d) 6.8\,eV} \\ \hline
\includegraphics[width=0.48\linewidth]{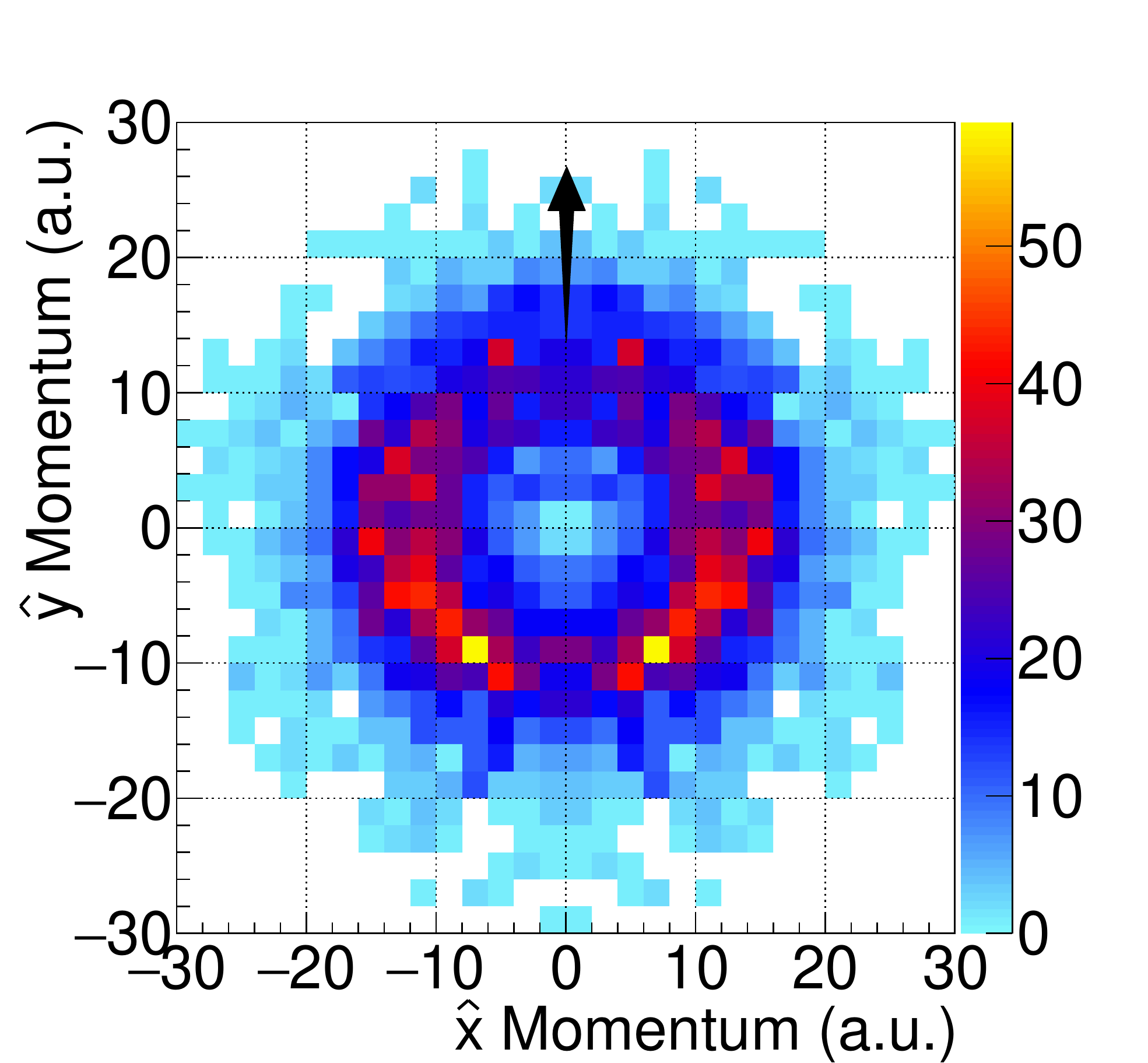} &
\includegraphics[width=0.46\linewidth]{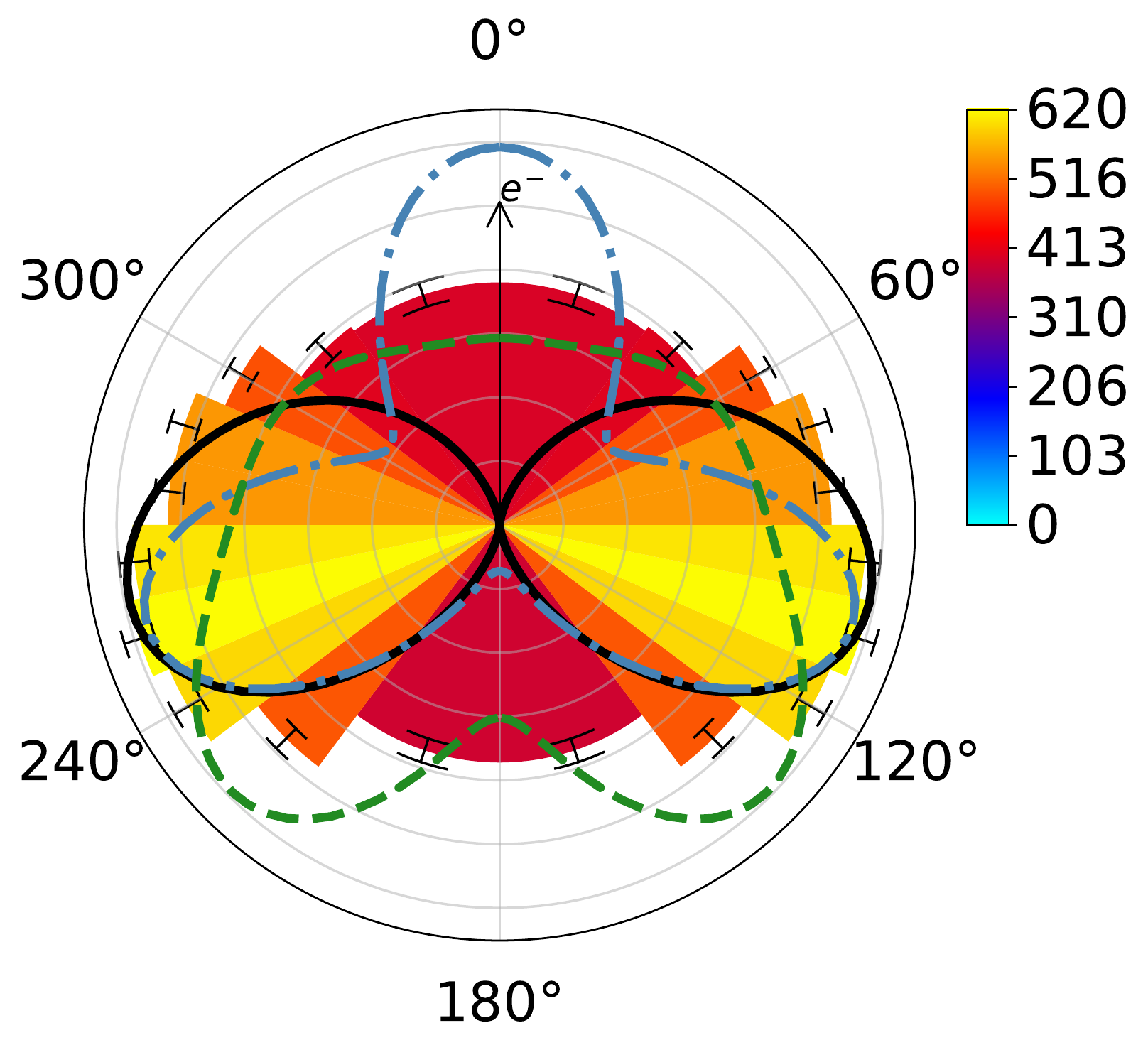} \\ 
\end{tabular}

\centering

\caption{\emph{Left}: Momentum images of NH$_2^-$ dissociation from DEA to formamide. Electron is incident in the +y-direction with energy (a) 5.3~eV, (b) 5.8~eV, (c) 6.3~eV, and (d) 6.8~eV. \emph{Right}: Histograms of dissociation angle of NH$_2^-$ anions from formamide with incident electron energy of (a) 5.3~eV, (b) 5.8~eV, (c) 6.3~eV, and (d) 6.8~eV, and incident at 0$^\circ$.  
The black line indicates the calculated angular distribution of NH$_2^-$ for DEA to the lowest $^2$A$''$ Feshbach resonance. The blue dot-dashed line represents the calculated angular distribution of NH$_2^-$ for the  $^2$A$'$ Feshbach resonance and the green dashed line shows the $^2$A$'$ with a 30 degree rotation of the recoil axis toward larger O-C-N angles.
}
\label{fig:NH2minus6p3}
\end{figure}

\paragraph{Momentum imaging.}
Momentum images of the NH$_2^-$ anion from DEA to formamide with incident electron energies between 5.3--6.8~eV are provided in Fig.~\ref{fig:NH2minus6p3}. 
These momentum images have a $\pi$/2-radian conical slice selection gate, which allows us to project the 3D momentum distribution in 2D with a uniform volume in momentum space. 
The compact momentum distribution of NH$_2^-$ was subject to small distortions in the imperfect spectrometer fields, which we have addressed in the calibration procedure, and by exploiting the symmetry about the incident electron axis (which is along the y-axis).

The momentum images and accompanying angular plots for NH$_{2}^{-}$ are shown in   Fig.~\ref{fig:NH2minus6p3}. We observe  that the distribution is isotropic at the lower incident electron energies, but builds up a maximum fragment yield at about 105$^\circ$ from the incident electron direction axis in the 6.8~eV data.
In addition to the experimental data, the right column of Fig.~\ref{fig:NH2minus6p3} shows the computed angular distributions for $^2$A$''$ (black continuous line) and  $^2$A$'$ (blue dot-dashed line) Feshbach  resonances, calculated under the assumption that the axial-recoil approximation applies - i.e. that the dissociating bond does not rotate during fragmentation. While the $^2$A$''$ resonance gives a predicted distribution that is consistent with the measured distributions that show the prevalence of a peak around $\sim$105$^\circ$ at the higher end of the 5.3--6.8~eV energy range, the calculated distribution for the $^2$A$'$ resonance is highly non-isotropic and bears little resemblance with the measured distributions at any energy. This is consistent with a breakdown of axial recoil for this resonance pointing to an internal geometric change of the transient resonance anion prior to breakup. In an attempt to model this, we have also included the angular distribution of the $^2$A$'$ Feshbach resonance for a modified dissociation axis, which is modeled on a 30$^\circ$ rotation of the C-N dissociation axis towards larger O-C-N angle (green dashed line). This gives a more isotropic angular distribution that better agrees with the measured values at the lower end of the electron energy range, where the $^2$A$'$ resonance presumably dominates.

\begin{figure}[h!]
    \centering
    \includegraphics[width=\linewidth]{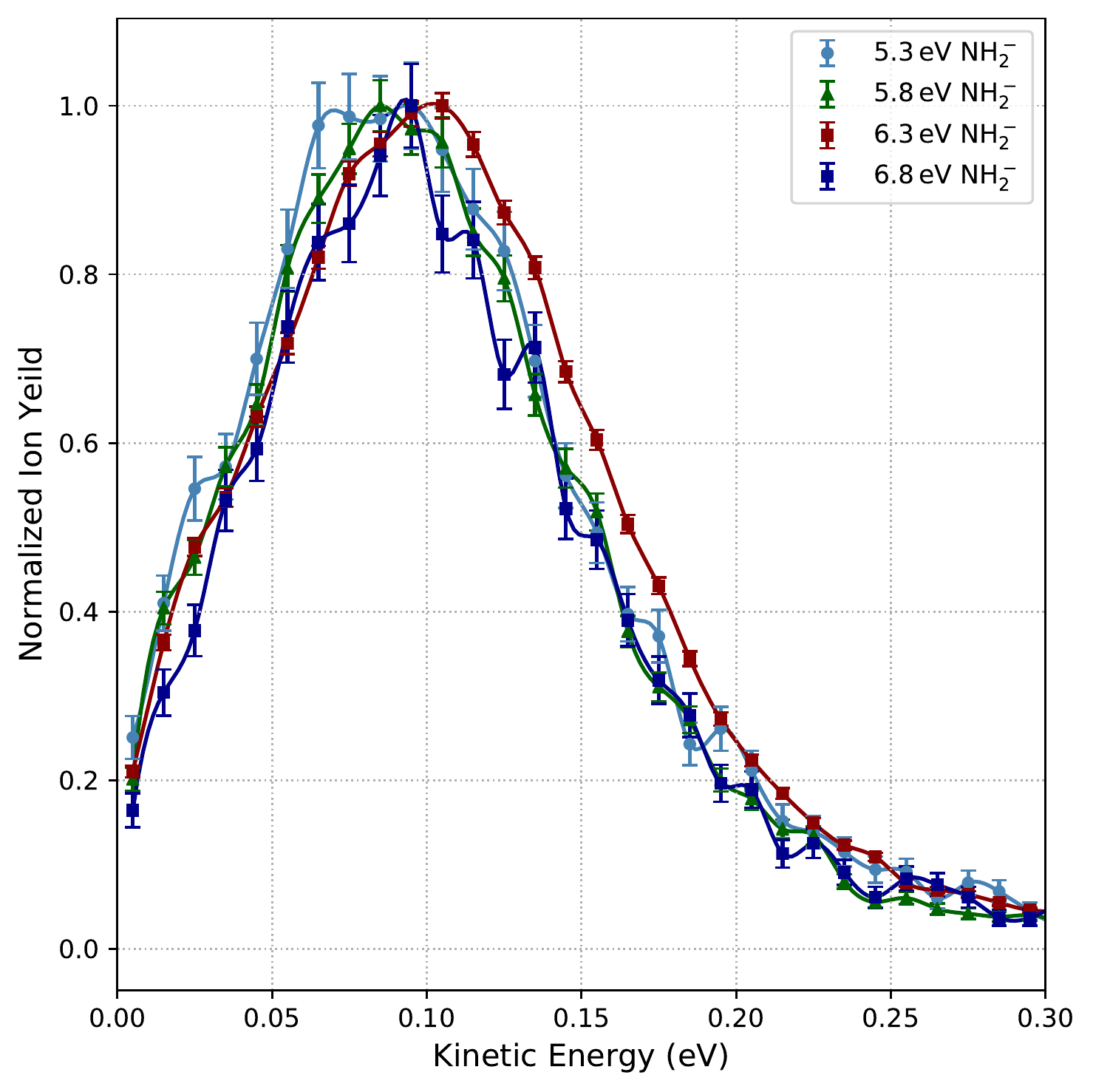}
    \caption{Kinetic energy spectra of NH$_2^{-}$ fragment from DEA to formamide from incident electron energies 5.3--6.8~eV. 
    }
    \label{fig:NH2KE}
\end{figure}

In Fig.~\ref{fig:NH2KE} we provide the kinetic energy distributions for the NH$_2^-$ fragment for each electron energy. 
We find that the maximum of the NH$_2^-$ distribution occurs at approximately $\sim$0.09~eV. The kinetic energy distribution does not change considerably with the incident energy over this range of incident energies. The peak at $\sim$0.09~eV indicates a three-body dissociation, and/or significant rovibrational excitation in the molecular fragments, producing NH$_{2}^{-}$ and H + CO or HCO in the three-body or two-body dissociation, respectively. For the limiting case of a prompt three-body dissociation into cold molecular fragments, the neutral H atom could be released with as much as 1.6--3.1~eV over this incident energy range. In the limiting case of two-body dissociation into rovibrationally hot molecular fragments, 1.6--3.1~eV is the available internal energy in the excited NH$_{2}^{-}$ and HCO fragments.

Previous mass-resolved anion fragment yield experiments~\cite{Hamann2011,Li2019} revealed two peaks producing NH$_{2}^{-}$ at 5.9~eV and 6.8~eV.  
Despite the very small momentum available to the detected NH$_2^-$ fragment, the subtle changes we find in the NH$_2^-$ momentum images and  angular distributions between 5.9~eV and 6.8~eV, coupled with the theoretical predictions of two Feshbach resonances in this energy range, support the notion of two overlapping resonances contributing to the NH$_2^-$ production. 
The 6.8~eV resonance is expected to have the dominant contribution at the higher incident electron energies due to the higher NH$_2^-$ yield previously observed at 6.8~eV, relative to the smaller peak that occurs on the low-energy shoulder around 5.9~eV~\cite{Hamann2011,Li2019}.


\subsection{O$^{-}$ resonance at 10.0 - 11.5~eV}\label{sec:O}

\paragraph{Pathways to formation.}
O$^-$ is one of the dominant fragments for incident electrons with energies between 10-11.5~eV~\cite{Hamann2011}. The formation of O$^{-}$ anions through DEA to formamide requires the cleaving of the C-O double bond with a bond dissociation enthalpy of 7.7~eV. There are three possible two-body breakups of formamide and several feasible three-body processes. We will provide the two-body mechanisms here.

The two-body fragmentation that produces O$^{-}$ along with neutral aminomethylene (HCNH$_2$) is given by:
$$
    e^{-} + \textrm{HCONH}_2 \to (\textrm{HCONH}_2)^{\ast -} \to \textrm{HCNH}_2 + \textrm{O}^{-} \, .
$$
From the thermodynamic data in Table~\ref{tab:tab1}, the threshold for this process is expected to be 6.2~eV. Additional two-body processes can be conceived by rearrangement of the hydrogen atoms in the neutral counterpart to O$^-$. This includes:
$$
    e^{-} + \textrm{HCONH}_2 \to (\textrm{HCONH}_2)^{\ast -} \to \textrm{H}_2\textrm{CNH} + \textrm{O}^{-} \, ,
$$
and
$$
    e^{-} + \textrm{HCONH}_2 \to (\textrm{HCONH}_2)^{\ast -} \to \textrm{CH}_3\textrm{N} + \textrm{O}^{-} \, .
$$
These processes have thresholds of 4.2~eV and 6.3~eV, respectively (based on the values in Table~\ref{tab:tab1}).

The three-body mechanism that produces O${^-}$, HCN and molecular hydrogen H$_2$, due to two N-H bonds and a C-O bond breaking, has a threshold of 4.4~eV, while the three-body mechanism producing O${^-}$, HCN and H$_2$, by C-H break, N-H break and C-O break, has a threshold of 5.0~eV. Consequently, these three-body processes may also play a role in the observed resonance between 10-11.5~eV.

\begin{figure}[b!]
\begin{tabular}{c c}
\multicolumn{2}{c}{(a) 10.0\,eV } \\ \hline 
\includegraphics[width=0.48\linewidth]{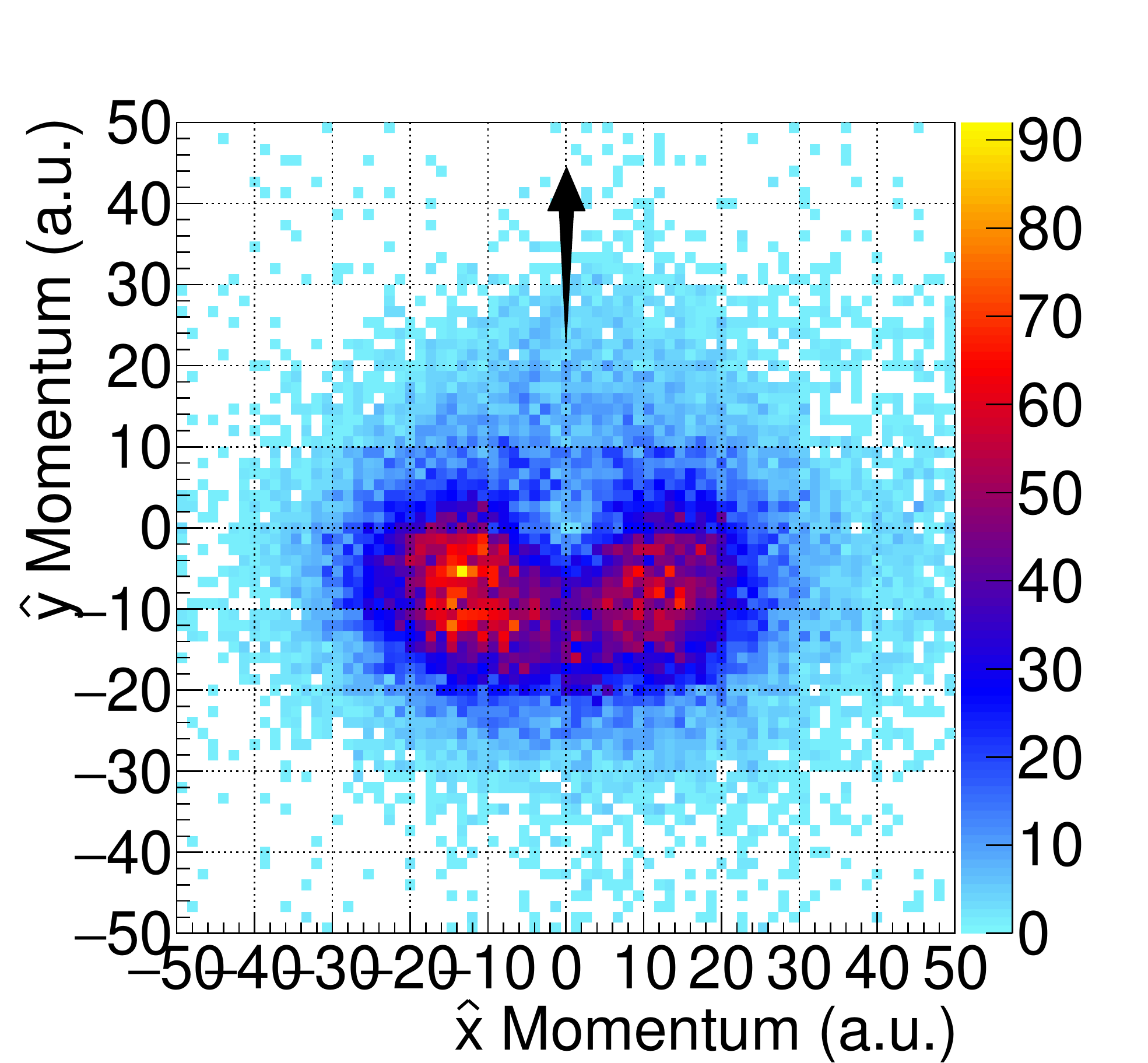} &
\includegraphics[width=0.46\linewidth]{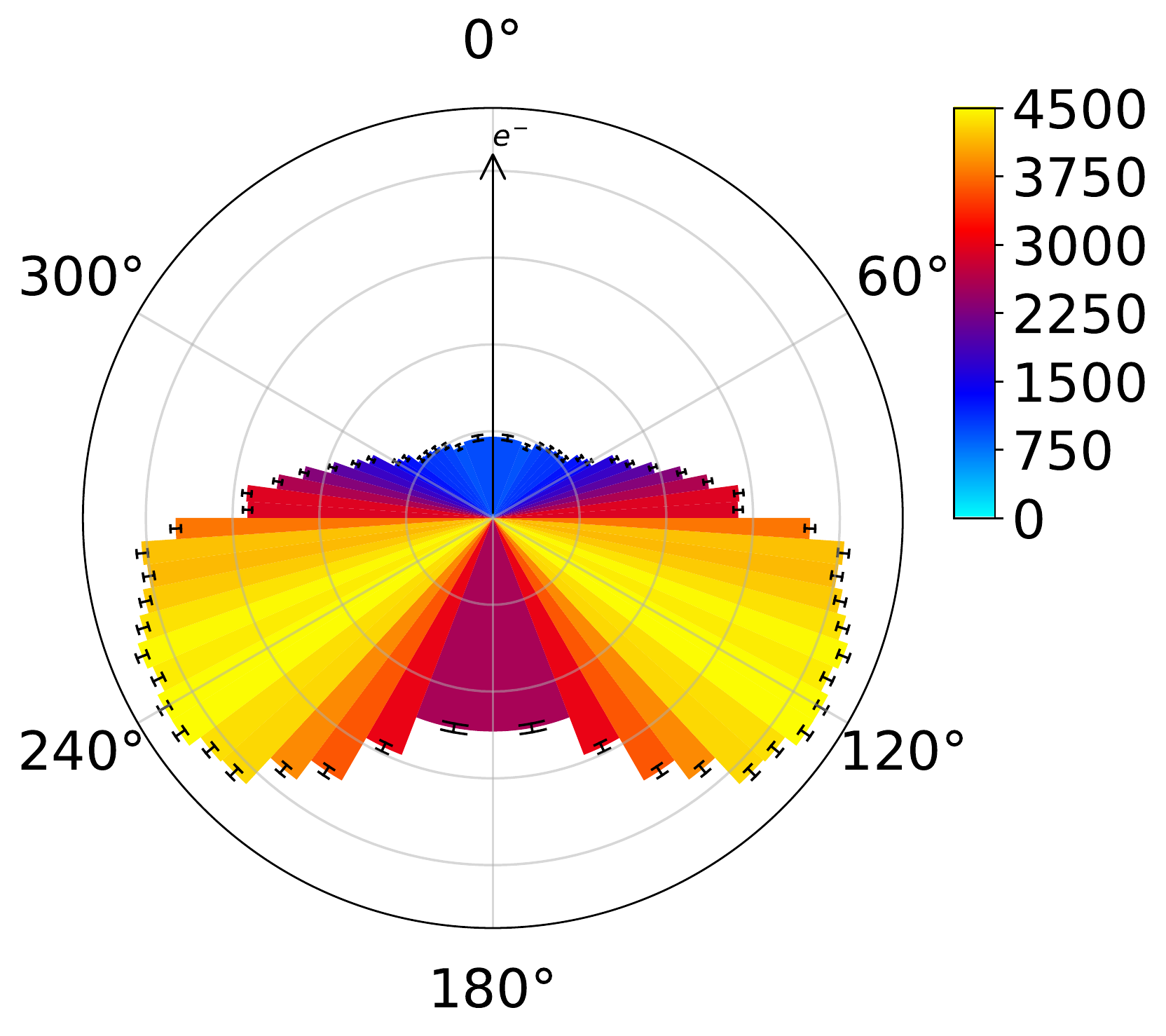} \\  \multicolumn{2}{c}{(b) 10.5\,eV} \\ \hline
\includegraphics[width=0.48\linewidth]{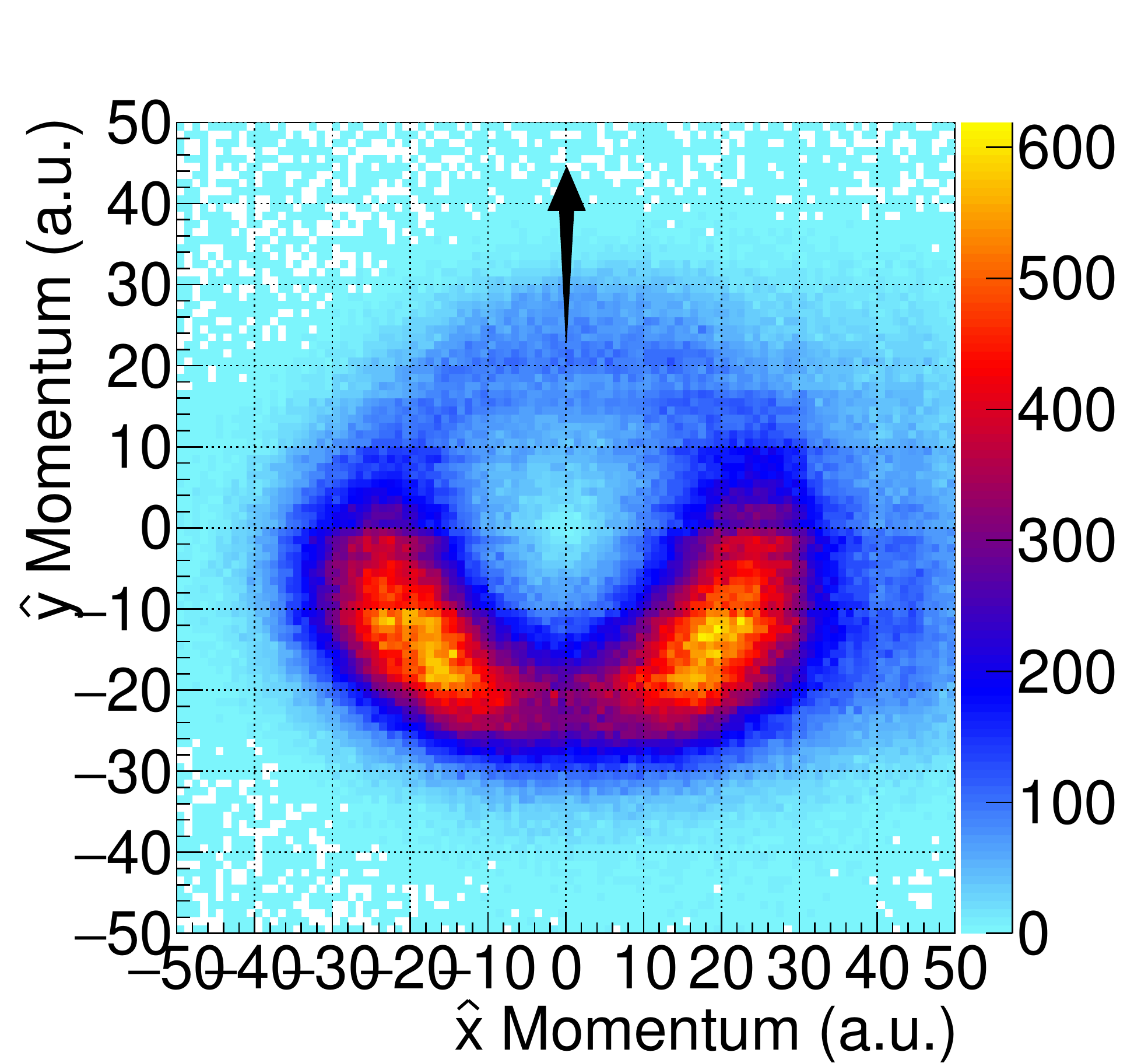} &
\includegraphics[width=0.46\linewidth]{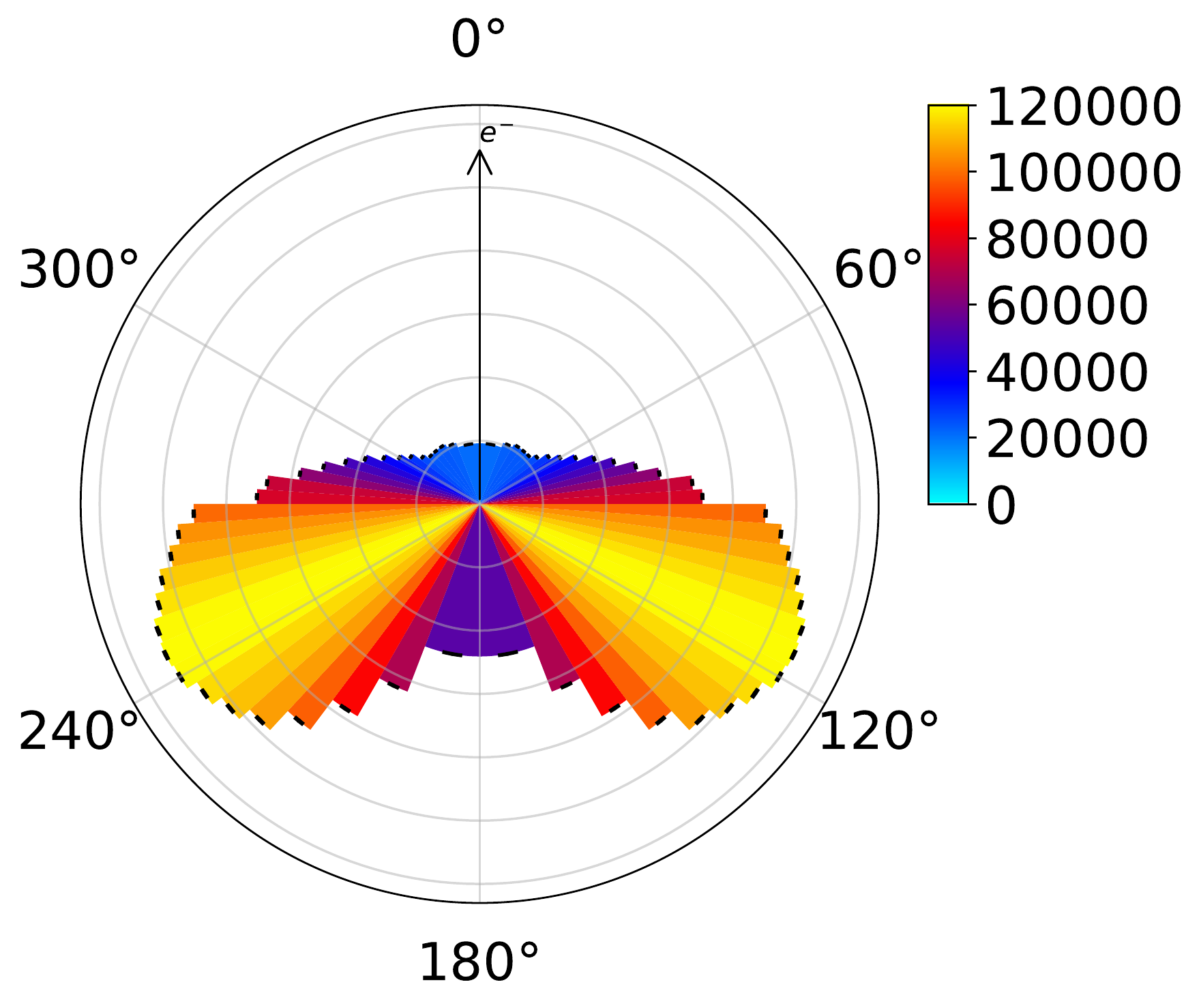} \\   \multicolumn{2}{c}{(c) 11.0\,eV} \\ \hline
\includegraphics[width=0.48\linewidth]{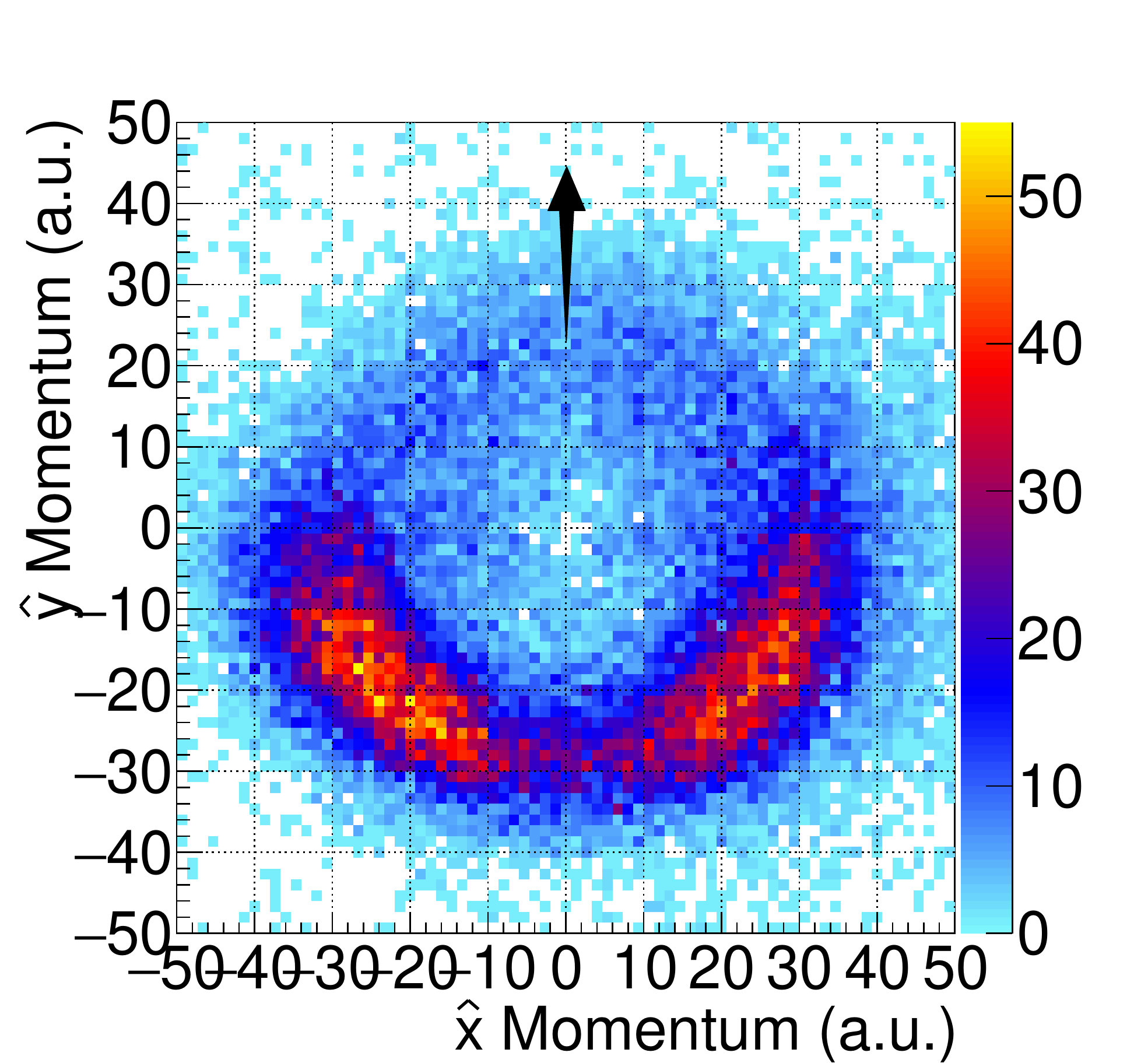} &
\includegraphics[width=0.46\linewidth]{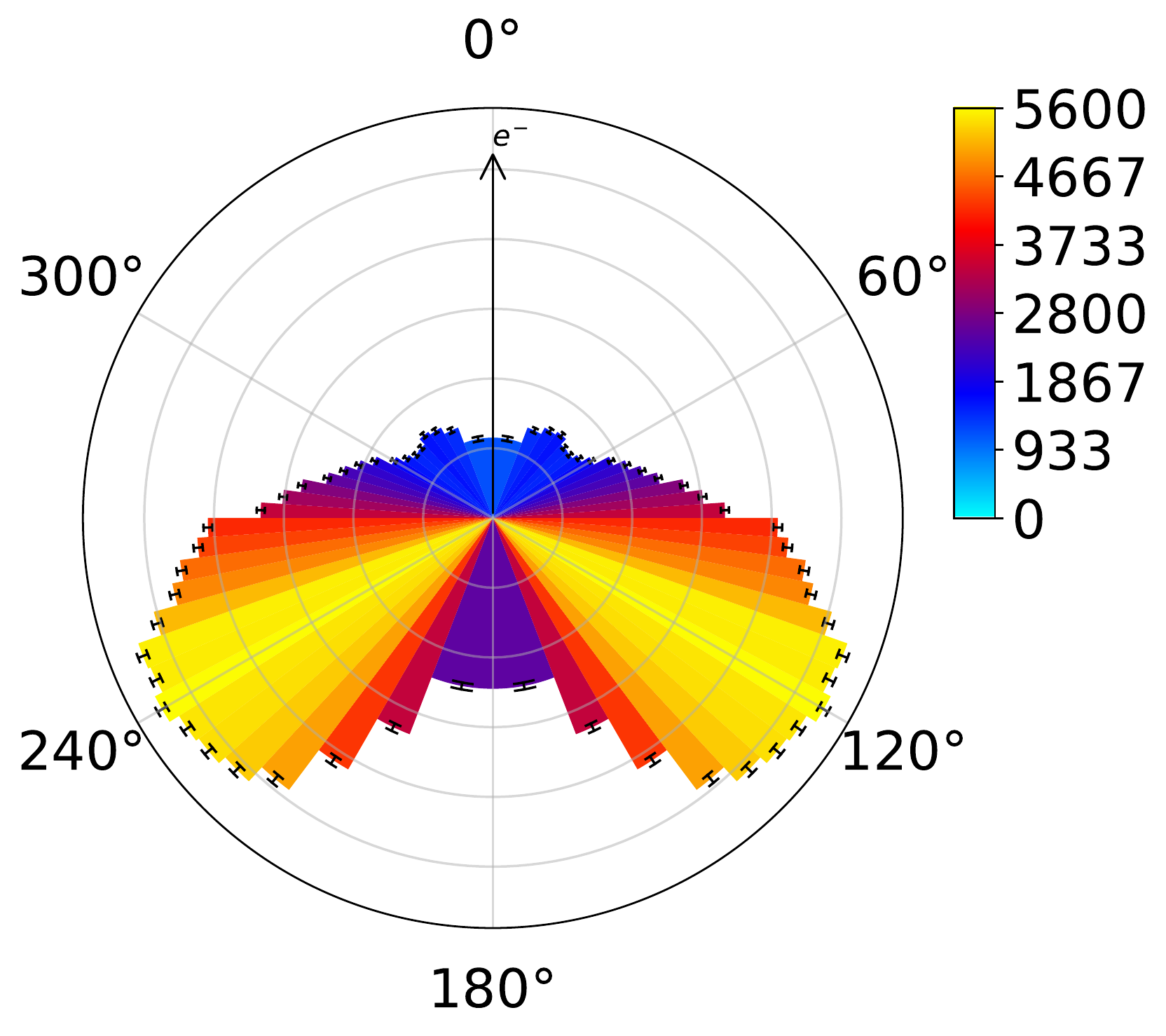} \\  \multicolumn{2}{c}{(d) 11.5\,eV} \\ \hline
\includegraphics[width=0.48\linewidth]{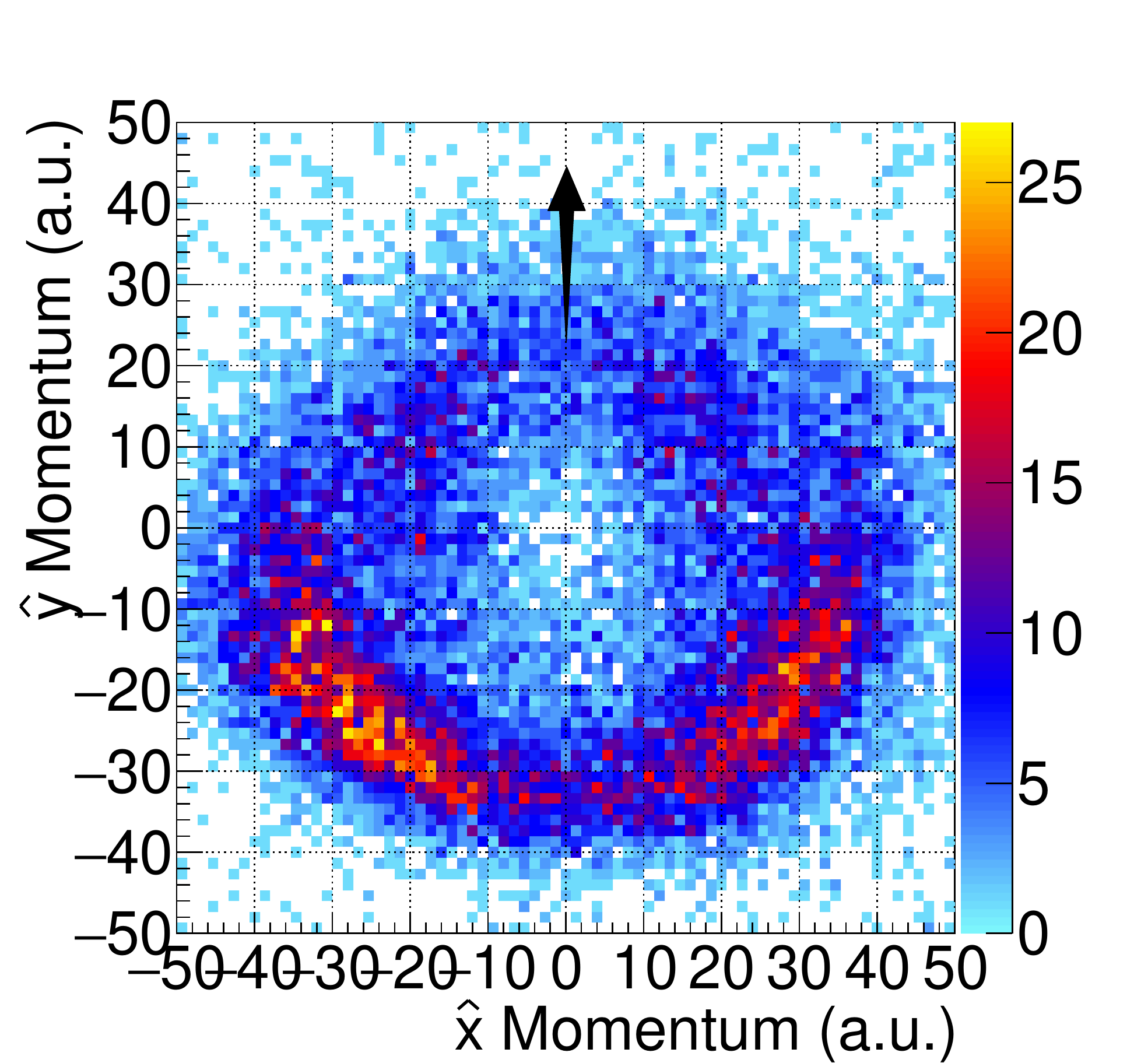} &
\includegraphics[width=0.46\linewidth]{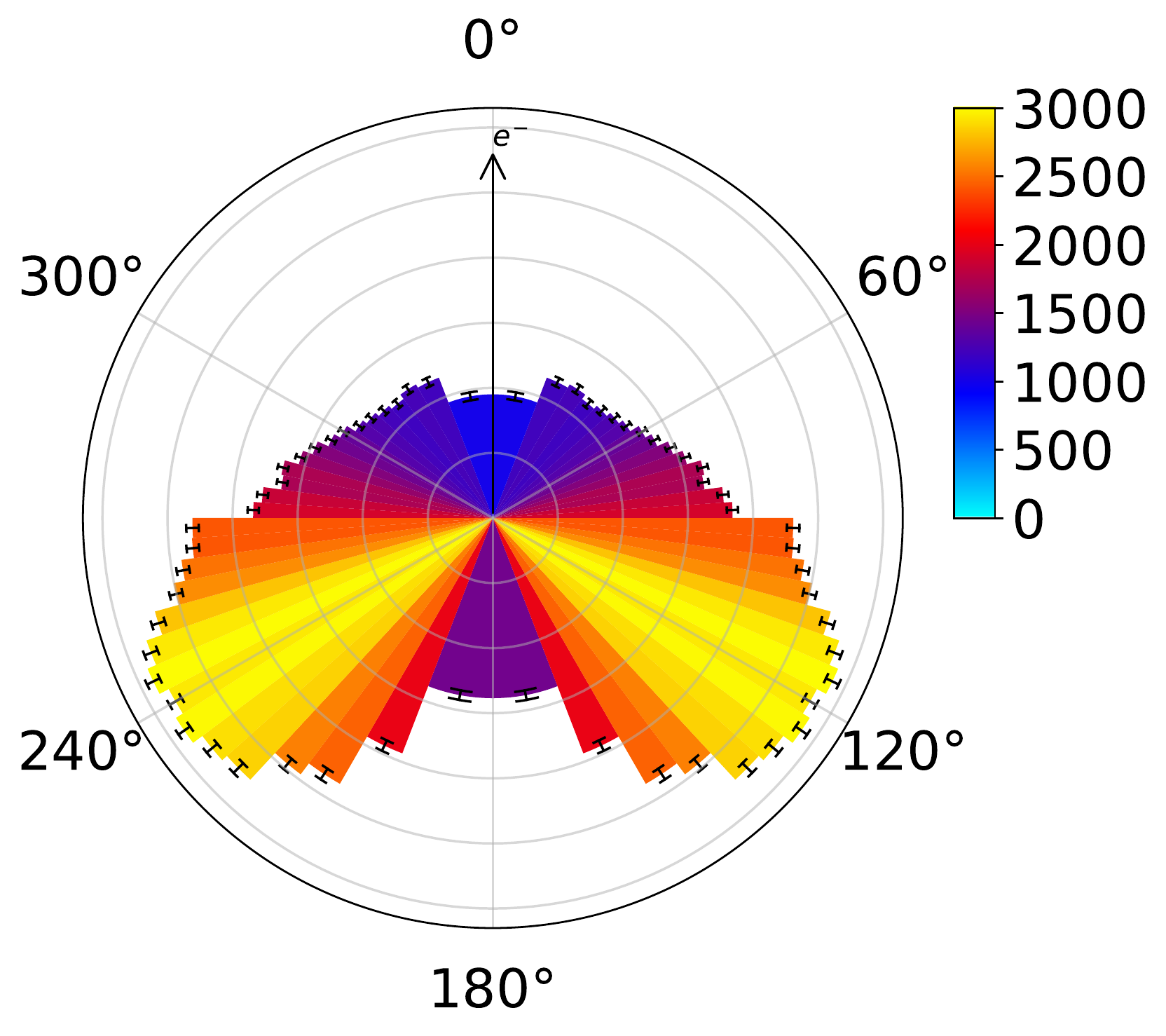} \\ 
\end{tabular}

\centering

\caption{\emph{Left}: Momentum images of O$^-$ dissociation from DEA to formamide. Electron is incident in the +y-direction with energy (a) 10.0~eV, (b) 10.5~eV, (c) 11.0~eV, and (d) 11.5~eV. \emph{Right}: Histograms of dissociation angle of O$^-$ anions from formamide with incident electron energy of (a) 10.0~eV, (b) 10.5~eV, (c) 11.0~eV, and (d) 11.5~eV in the incident at 0$^\circ$. 
}
\label{fig:Omom}
\end{figure}

\paragraph{Momentum imaging.}

The measured momentum distributions for the O$^-$ anion fragment from DEA to formamide is provided in Fig.~\ref{fig:Omom}. Here we incorporated a $\pi$/2-radian selection gate on the 3D momentum sphere. Note that our analysis of O$^-$ fragments did not impose cylindrical symmetry. This leads to momentum distributions that are slightly asymmetric about the incident electron direction axis, possibly due to minor imperfections in the electric fields within the spectrometer, small variations in detection efficiency across the face of the detector, and statistical uncertainties.

The O$^-$ momentum is sharply peaked at $\sim$120$^\circ$ from the incident electron direction. The angular distributions of the O$^-$ anion fragment provided in Fig.~\ref{fig:Omom} (right column) explicate this fact. 
This suggests that the O$^-$ fragments are ejected promptly along the C$\to$O bond axis with little or no rotation of the C-O bond. The 120$^\circ$ angle between the dissociating C-O bond and the incident electron is consistent with a high electron attachment probability for incident electrons along the C$\to$N direction. The OCN bond angle was calculated previously~\cite{fogarasiHighLevelElectronCorrelation1997} for the equilibrium geometry of neutral formamide to be $\sim$125$^\circ$. 

\begin{figure}[b!]
    \centering
    \includegraphics[width=\linewidth]{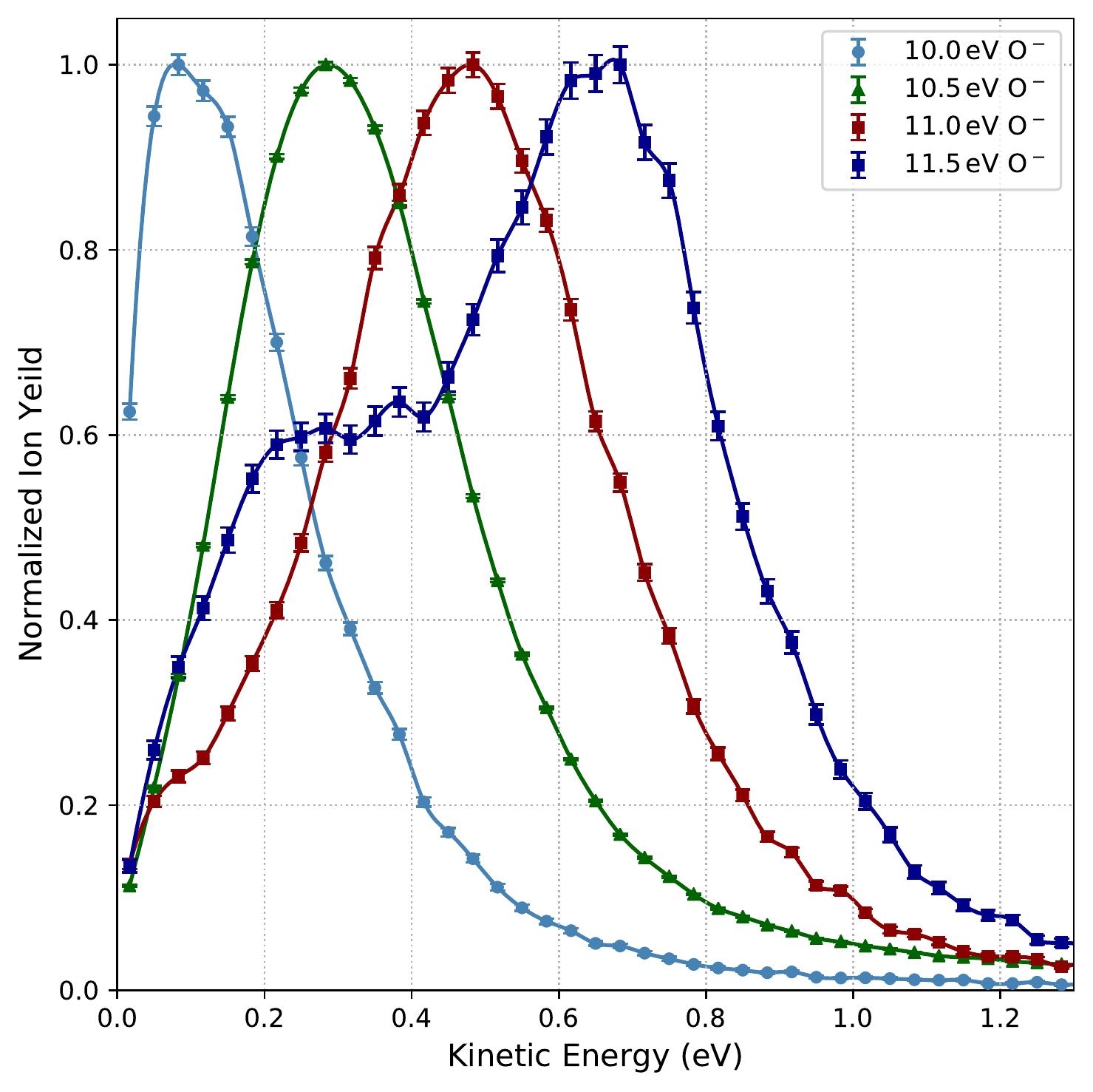}
    \caption{Kinetic energy spectra of O$^-$ fragment produced from DEA to formamide at incident electron energies 10.0-11.5~eV. The lower-energy peak of the double peak structure for the 11.5~eV curve is attributed to O$^-$ from gaseous water within the interaction region. 
    }
    \label{fig:OKE}
\end{figure}

The peak magnitude of the momentum distributions increases with the incident electron energy (from $\sim$10~a.u. at 10.0~eV to $\sim$25~a.u. at 11.5~eV)
The kinetic energy spectra of O$^-$ anion for 10-11.5~eV incident electrons are displayed in Fig.~\ref{fig:OKE}. The maxima of these kinetic energy spectra increase approximately linearly with the incident electron energy. In addition to the formamide target gas, background gas in the target region includes a small but significant presence of water. As the incident electron energy increases, so does the production of O$^-$ anions from DEA to water~\cite{Fedor2006}. While the impact of water background on the O$^-$ signal is small, we attribute the double peak feature in the kinetic energy distribution for the highest present incident electron energy of 11.5~eV to water contamination in the interaction region. Additionally, the forward component (between 270$^\circ$ and 90$^\circ$) of the angular distribution starts to increase at 11.5~eV, which is also indicative of water contamination~\cite{adaniyaImagingMolecularDynamics2009,haxtonObservationDynamicsLeading2011}. For the limiting case of two-body dissociation, the internal energy in the HCNH$_2$ fragment, which may subsequently isomerize or undergo a secondary dissociation, is 3.6~eV to 4.2~eV.

\subsection{H$^{-}$ resonance at 10.0 - 11.5~eV}

The production of H$^{-}$ anions from DEA to formamide is known to have a peak yield at an incident electron energy of 6.5~eV along with a less dominant peak at 10.5~eV~\cite{Hamann2011}. In the present experiments the substantial background of H$^-$, due to the small but significant water contamination in the interaction region, produced ambiguous results around 6.5~eV incident energies, so they are not presented here. At higher incident energies the much smaller DEA cross section for water reduces the H$^-$ contamination by $\sim$2 orders of magnitude, so the present results for H$^-$ from formamide in the 10.0~eV to 11.5~eV range of incident electron energies are essentially background-free.

\paragraph{Pathways to formation}
The process of DEA to formamide resulting in the dissociation of H$^{-}$ via cleaving of an N-H bond proceeds as:
$$
    e^{-} + \textrm{HCONH}_2 \to (\textrm{HCONH}_2)^{\ast -} \to \textrm{HCONH} + \textrm{H}^{-} \, .
$$
Considering the bond dissociation enthalpy of the N-H bond (4.71~eV) and the electron affinity of atomic hydrogen (0.75~eV), we find that the threshold for this production mechanism is 3.9~eV, well below the incident electron energy for the observed resonance.

\begin{figure}[t!]
\begin{tabular}{c c}
\multicolumn{2}{c}{(a) 10.5~eV } \\ \hline 
\includegraphics[width=0.48\linewidth]{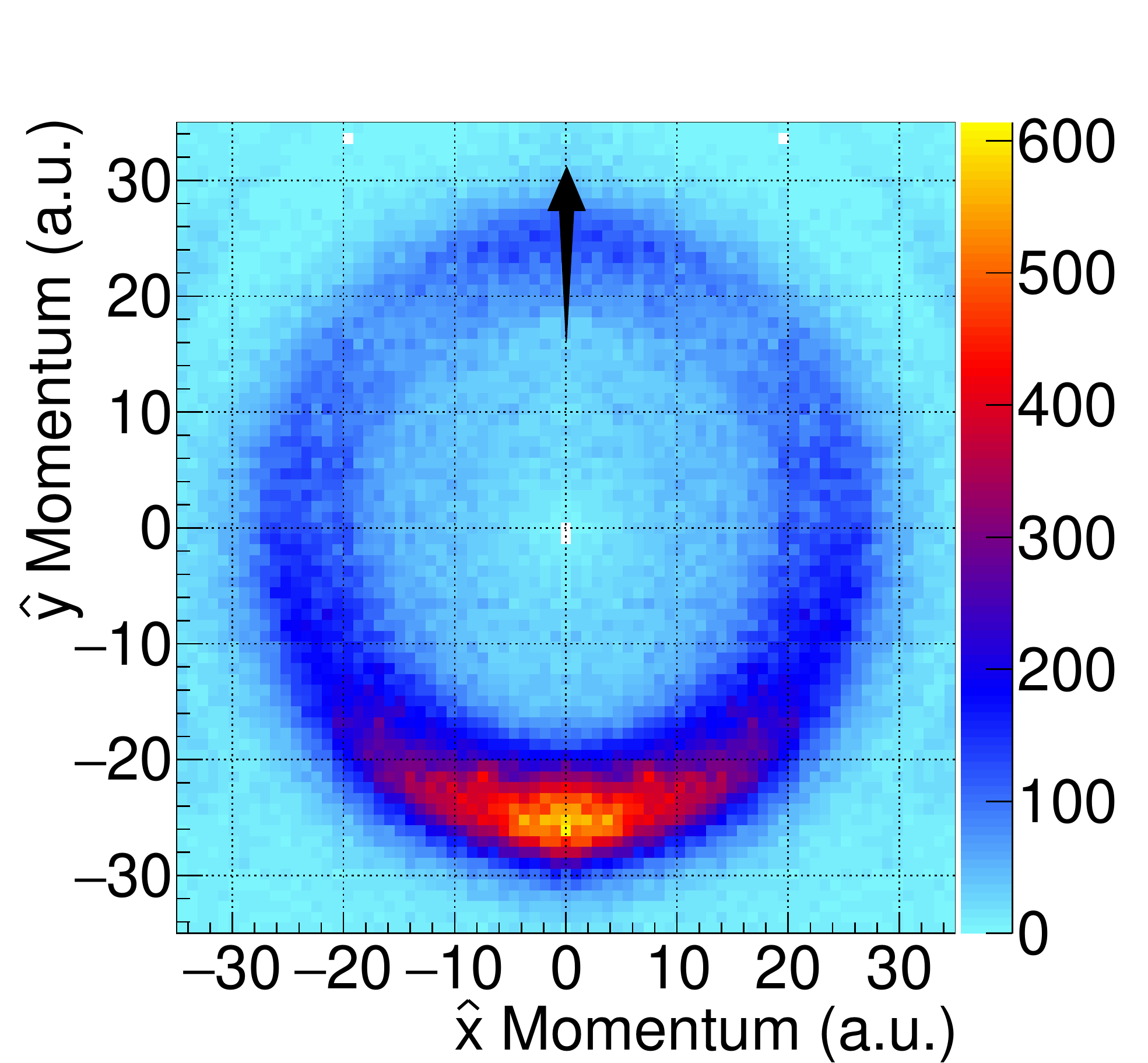} &
\includegraphics[width=0.46\linewidth]{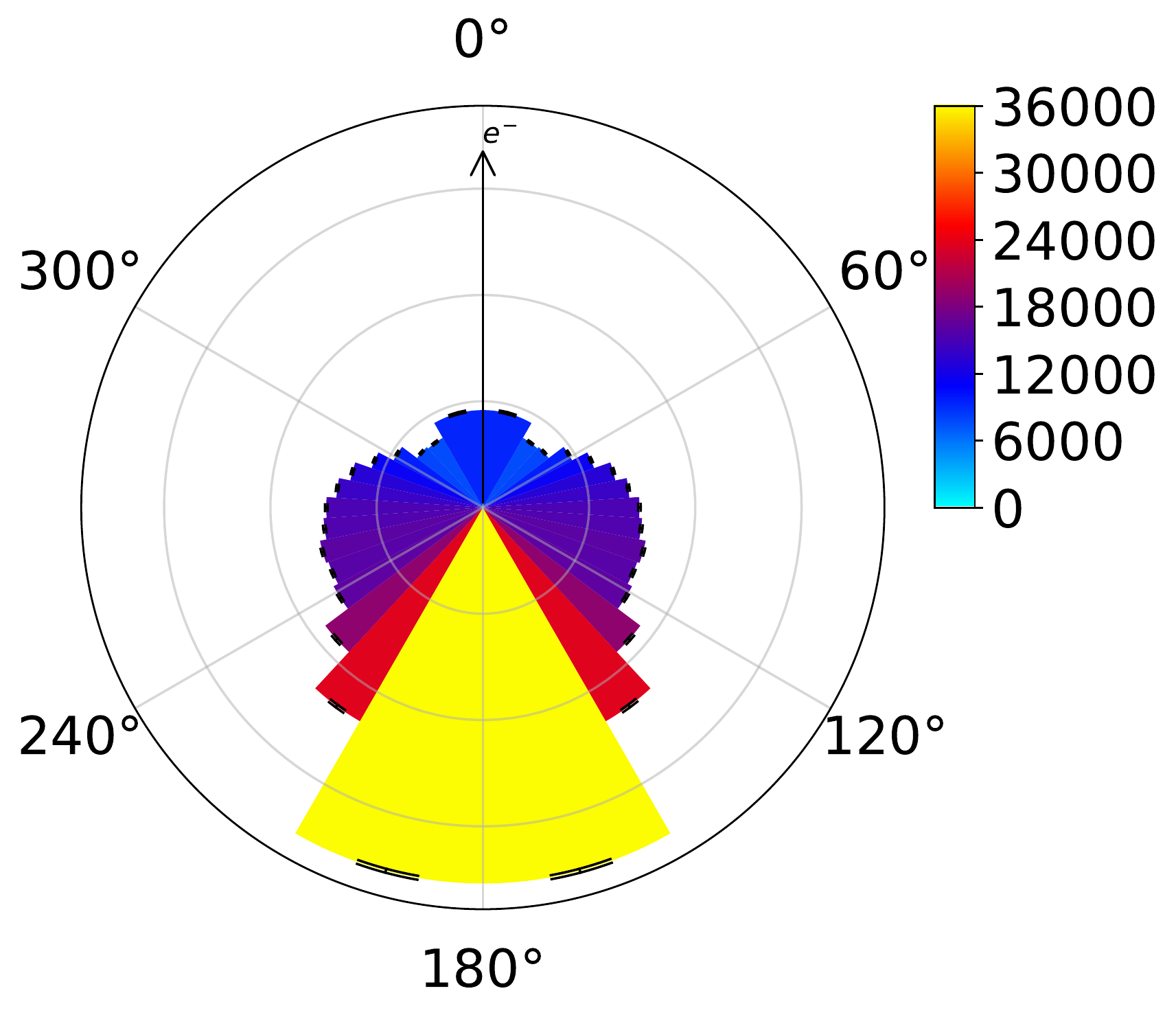} \\  \multicolumn{2}{c}{(b) 11.0\,eV} \\ \hline
\includegraphics[width=0.48\linewidth]{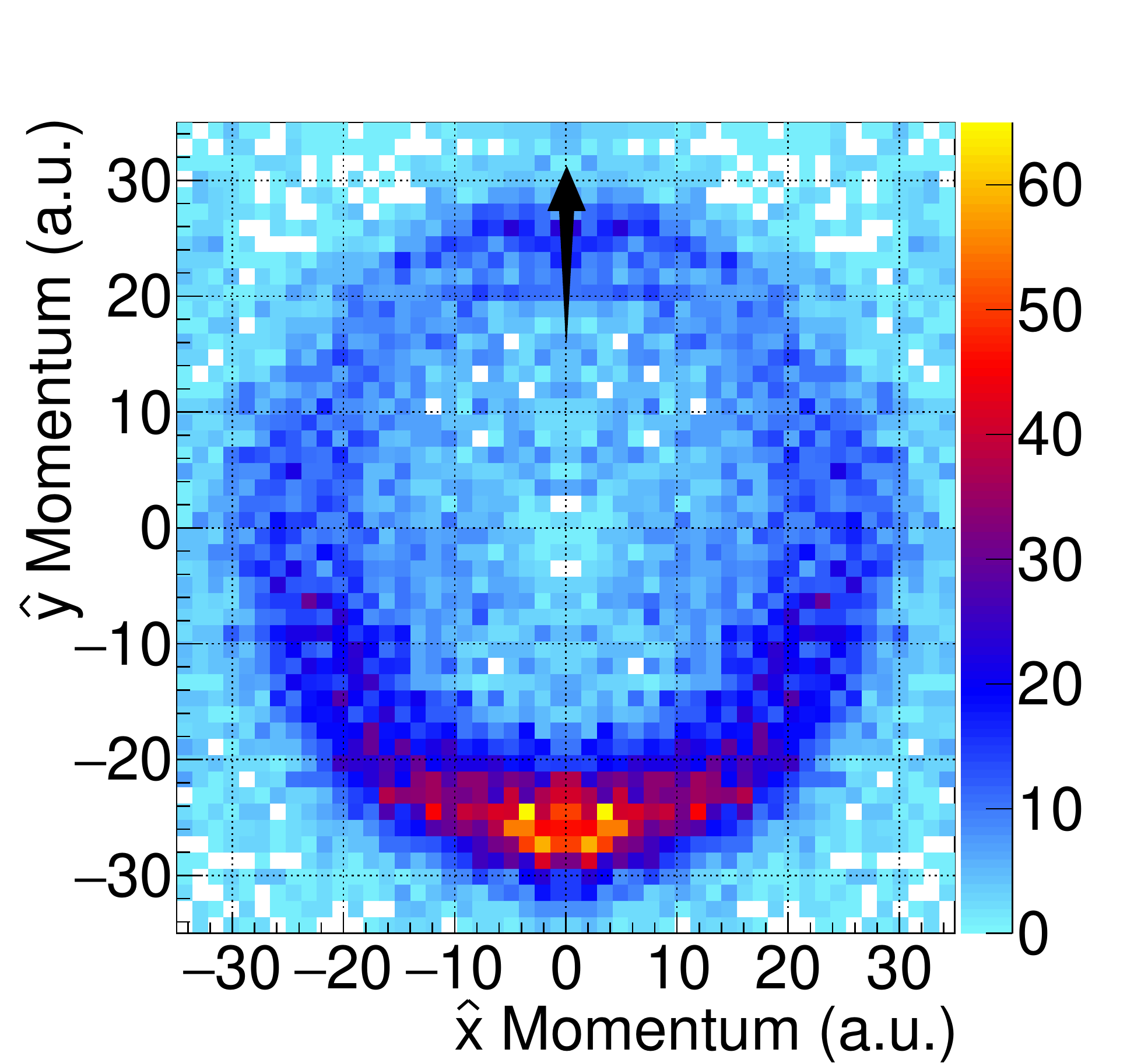} &
\includegraphics[width=0.46\linewidth]{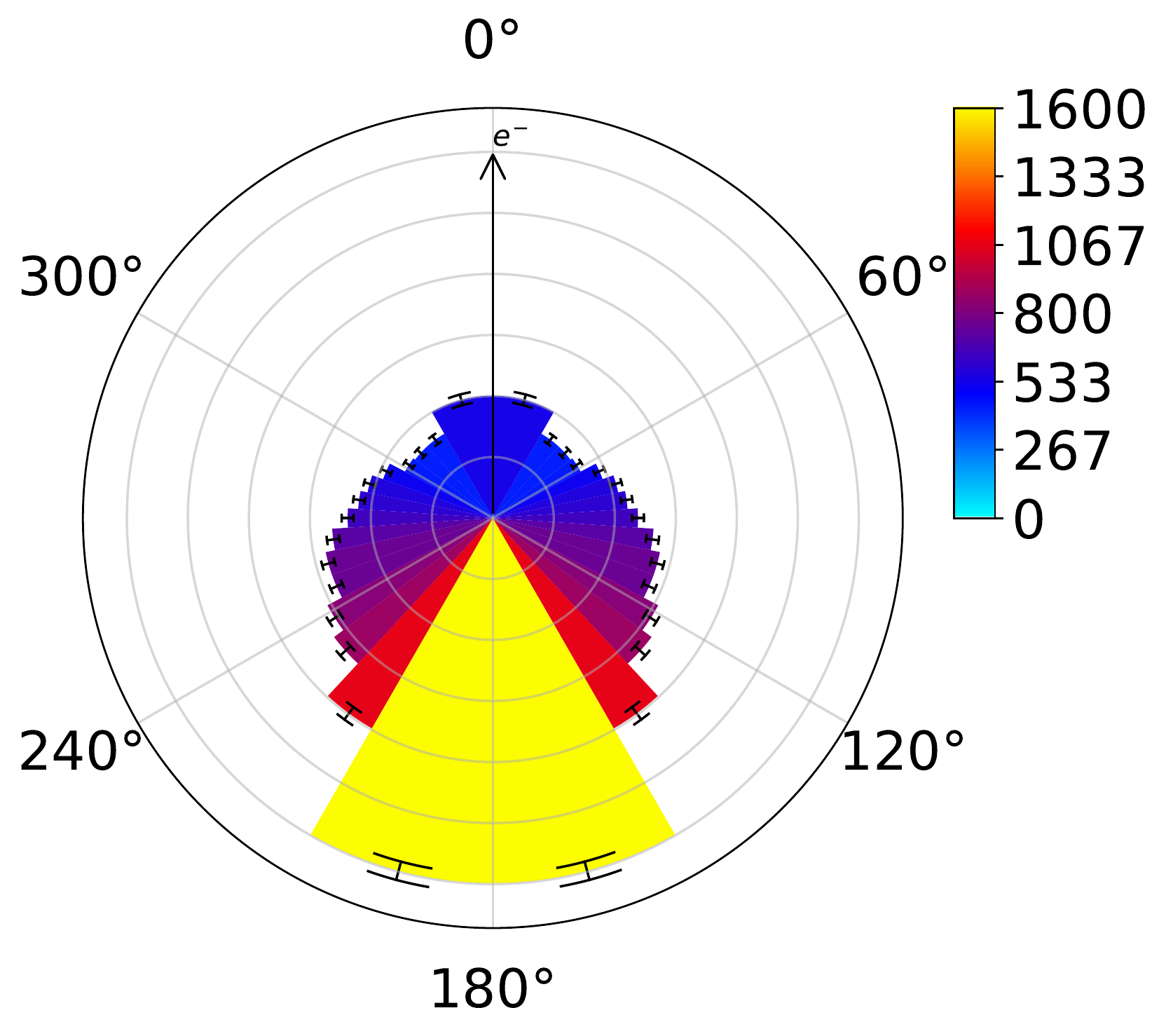} \\   \multicolumn{2}{c}{(c) 11.5\,eV} \\ \hline
\includegraphics[width=0.48\linewidth]{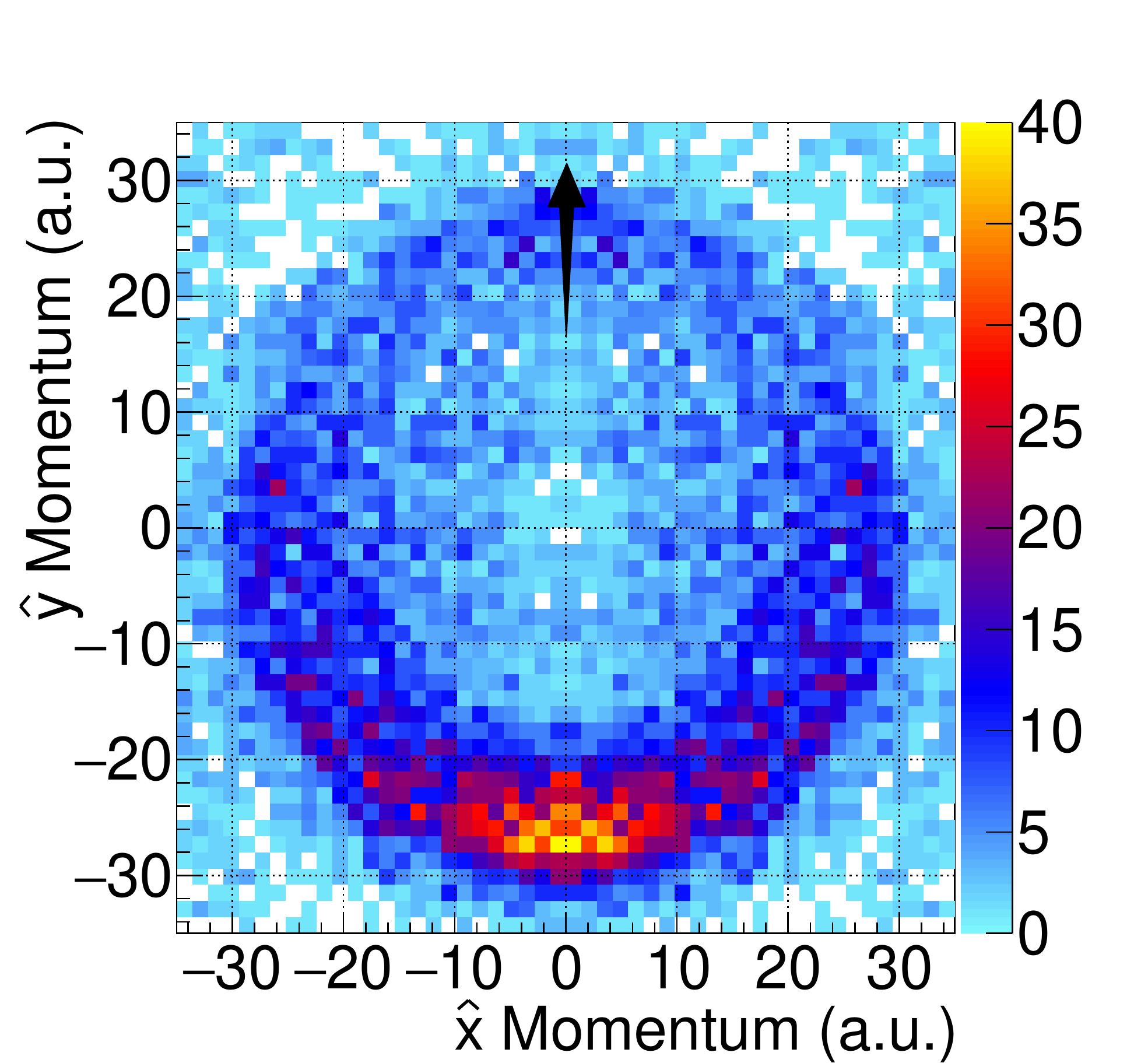} &
\includegraphics[width=0.46\linewidth]{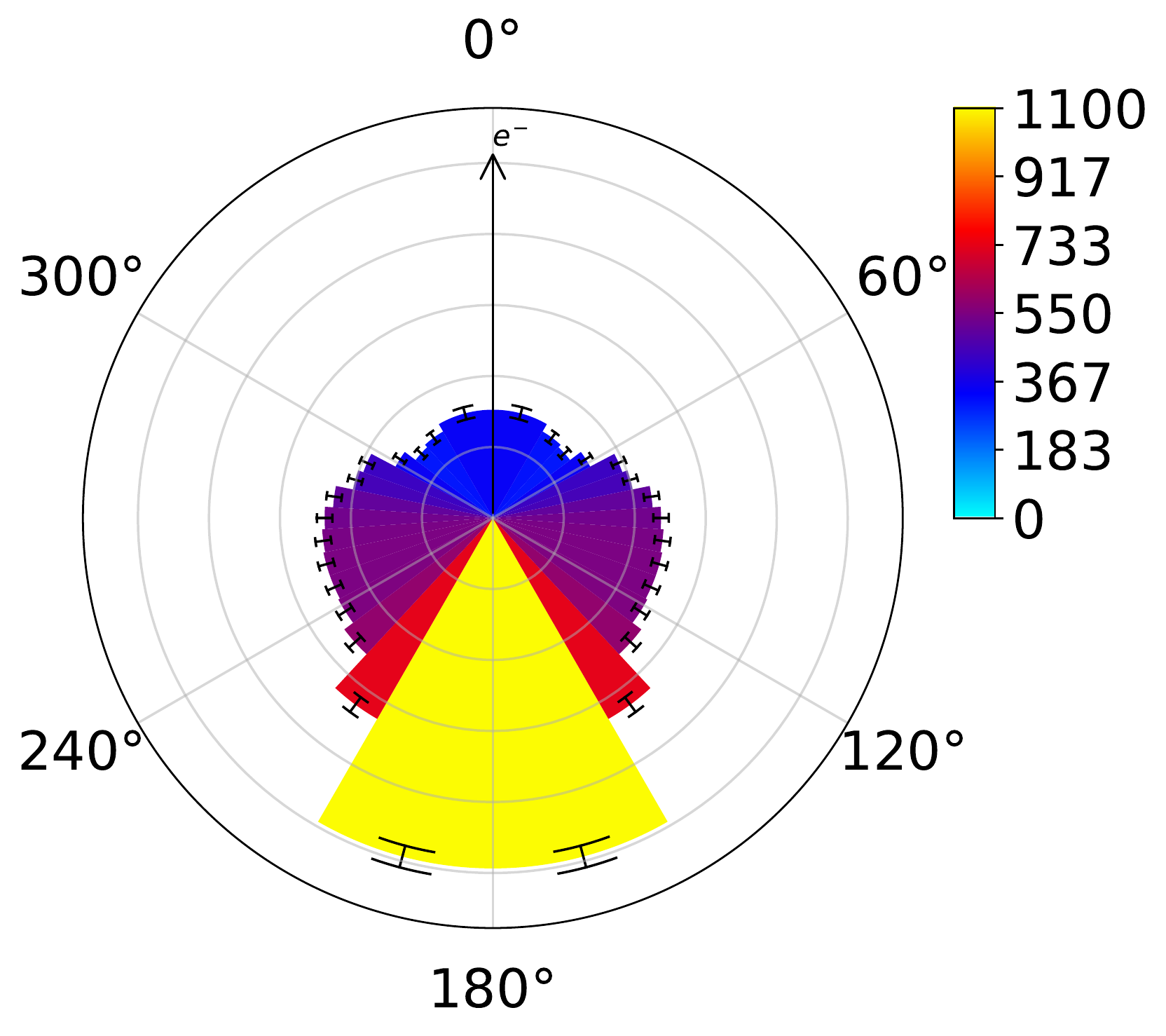} \\  
\end{tabular}

\centering

\caption{\emph{Left}: Momentum images of H$^-$ dissociation from DEA to formamide. Electron is incident in the +y-direction with energy (a) 10.5~eV, (b) 11.0~eV, and (c) 11.5~eV. \emph{Right}: Histograms of dissociation angle of H$^-$ anions from formamide with incident electron energy of (a) 10.5~eV, (b) 11.0~eV, and (c) 11.5~eV in the direction of 0$^\circ$. 
}
\label{fig:Hmom}
\end{figure}

\begin{figure}
    \centering
    \includegraphics[width=\linewidth]{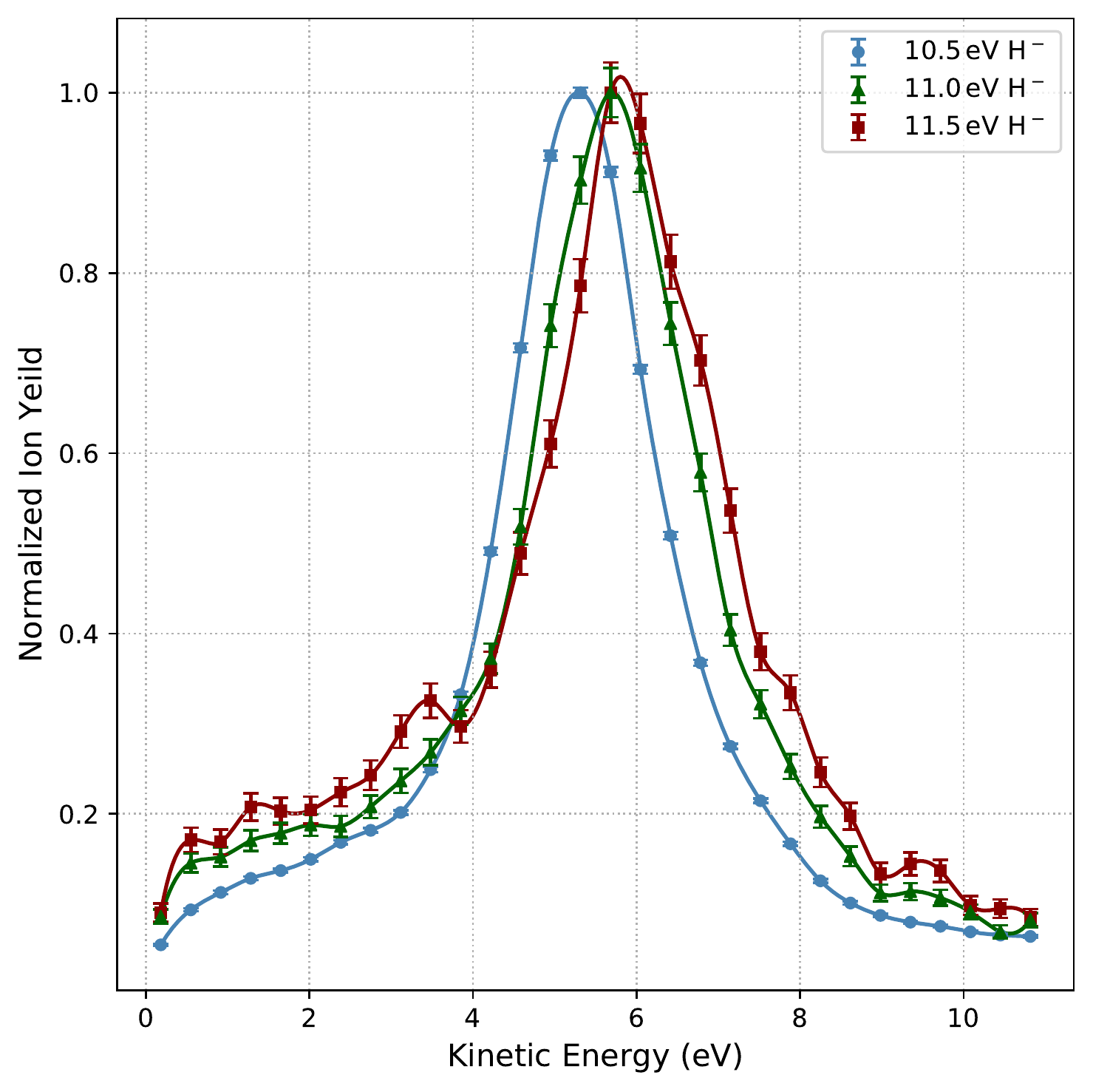}
    \caption{Kinetic energy spectra of H$^{-}$ fragment from DEA to formamide at incident electron energies 10.5--11.5~eV. 
    }
    \label{fig:HKE}
\end{figure}

The C-H bond cleaving mechanism may proceed in two fashions. The first of which results in the production of a hydrogen anion and a single neutral counterpart:
$$
    e^{-} + \textrm{HCONH}_2 \to (\textrm{HCONH}_2)^{\ast -} \to \textrm{CONH}_2 + \textrm{H}^{-} \, .
$$

Again, considering the electron affinity of atomic hydrogen and using the bond dissociation enthalpy for C-H from acetaldehyde (H$_3$CCHO), we find a threshold energy of 3.2~eV. The neutral CONH$_2$ in the reaction above may dissociate further into neutral CO and NH$_2$ molecular fragments, resulting in a third possible production mechanism for H$^{-}$ from formamide:
$$
    e^{-} + \textrm{HCONH}_2 \to (\textrm{HCONH}_2)^{\ast -} \to \textrm{CO} + \textrm{NH}_2 + \textrm{H}^{-}\, .
$$
Considering the standard heat of formation formamide, H$^{-}$, CO, and NH$_2$ from Table~\ref{tab:tab1}, this process has a threshold of 4.3~eV. Thus, both C-H cleaving formation processes are achievable at these incident electron energies.  While the formation at H$^-$ at the C site is possible, Hamann et al. \cite{Hamann2011} compared DEA on formamide to all its deuterated derivatives and their findings suggest that DEA at these higher incident electron energies results in H$^-$ due to N-H break without significant contributions of H$^-$ due to a C-H break.

\paragraph{Momentum imaging}
The momentum images for the H$^-$ fragmentation channel are shown in Fig. \ref{fig:Hmom} along with the respective angular distributions. Here we incorporated a $\pi/2$-radian selection gate on the 3D momentum sphere. In this fragmentation channel we observe that the H$^-$ anion strongly prefers to be emitted $\sim$180$^\circ$ relative to the incident electron. Smaller features are also visible: there is a shoulder at $\sim$110$^\circ$ and a small peak in the forward direction at $\sim$0$^\circ$. This angular distribution appears unchanged over the range of 10.5--11.5~eV. As for the O$^-$ angular distributions, the sharp structures and high kinetic energy indicate that the dissociation is prompt, and that little rotation of the dissociation axis occurs in the transient formamide anion prior to dissociation.  
The angular distribution is consistent with a high probability for electron attachment in the molecular frame of formamide along one H$\to$N direction, and H$^-$ loss primarily from the same bond. It is conceivable that H$^-$ loss may occur, with a lower yield, from the other N-H bond. In the equilibrium geometry of formamide, the H-N-H bond angle was previously calculated to be $\sim$120$^\circ$~\cite{fogarasiHighLevelElectronCorrelation1997}, which could be a possible explanation of  the shoulder at $\sim$110$^\circ$. 
The kinetic energy spectra of H$^-$ are shown in Fig. \ref{fig:HKE} and exhibit an increase in kinetic energy as the energy of the incident electrons increases. For the case of two-body dissociation, we can expect 1.2--1.7~eV to be available as internal energy in the HCNH fragment, which may subsequently isomerize or undergo a secondary dissociation.

\section{\label{sec:conclusions}Conclusions}
We have presented the results of anion 3D momentum measurements of DEA to formamide, leading to H$^-$, O$^-$, and NH$_2^-$ fragments in two energy regions, 5.3~eV to 6.8~eV and 10.0~eV to 11.5~eV. In the lower energy region, where two resonances were reported  previously~\cite{Hamann2011,Li2019}, the very low kinetic energy distribution of NH$_2^-$ does not change significantly between 5.3~eV and 6.8~eV. However, the angular distributions indicate a small but significant increase in anisotropy above 5.8~eV. \textit{Ab initio} electron scattering calculations of the electron attachment probability in the molecular frame for two Feshbach resonances offer insights into the character of the observed resonances. 
The qualitative agreement between the measured and calculated NH$_2^-$ angular distributions for the 6.3\,eV and 6.8\,eV measurements are consistent with the dominant DEA process producing NH$_2^-$ being due to electron attachment to the $^2$A$''$ Feshbach resonance. 
The calculated $^2$A$'$ Feshbach resonance angular distribution is a poor match for the lower two experimental distributions at 5.3~eV and 5.8~eV. 
The $^2$A$'$ angular distribution for opening O-C-N angles is much more isotropic, and thus it is more qualitatively consistent with the present experimental results. In summary, we find that the two resonances that dissociate by C-N break to form NH$_2^-$ are not necessarily dipole supported resonances, as recently reported~\cite{Li2019}. The present experimental measurements and {\it ab initio} electron scattering calculations suggest that these are in fact $^2$A$''$ and $^2$A$'$ Feshbach resonances with principle configurations (...)(2a$''$)$^1$ ($\sigma^*$)$^2$ and (...)(10a$'$)$^1$ ($\sigma^*$)$^2$, respectively. 

The anion resonances in the 10.0~eV--11.5~eV range are above the ionization threshold~\cite{ballardHePhotoelectronStudies1987} of formamide. Electron scattering calculations for resonances this high in the electronic continuum are highly challenging and beyond the scope of this work. Nevertheless, the sharp structures in the measured kinetic energy and angular distributions for the O$^-$ and H$^-$ fragments provide information on the possible electron attachment and dissociation mechanisms. Within the assumption that the dissociation axis does not rotate significantly prior to dissociation, the angular distributions for O$^-$ and H$^-$ indicate that each product may be formed from a distinct resonance, and that each of these two resonances has a distinct molecular-frame electron attachment probability.


\section*{Acknowledgements}
Work performed at the University of California Lawrence Berkeley National Laboratory was supported by the U.S. Department of Energy (DOE), Office of Science (SC), Division of Chemical Sciences of the Office of Basic Energy Sciences under Contract DE-AC02-05CH11231, and by the DOE SC, Office of Workforce Development for Teachers and Scientists, Office of Science Graduate Student Research (SCGSR) program. The SCGSR program is administered by the Oak Ridge Institute for Science and Education for the DOE under contract number DE‐SC0014664. Work performed at the University of Nevada, Reno was supported by the National Science Foundation Grant No. NSF-PHY-1807017. G.P. acknowledges support from the McNair Scholars Program at the University of Nevada, Reno. We are indebted to the RoentDek Company for long-term support with detector hardware and software.
\bibliography{ref.bib}

\end{document}